\documentclass[letterpaper,aps,prd,twocolumn,tightenlines,preprintnumbers,nofootinbib,showkeys,superscriptaddress,longbibliography]{revtex4-1}

\usepackage{lipsum} 

\usepackage{XCharter}
\usepackage[T1]{fontenc}
\usepackage{bm}
\usepackage{bbold}
\usepackage{pifont}
\usepackage{rotating}
\usepackage{fullpage}
\usepackage{amsfonts,shorthand}
\usepackage{amsmath}
\usepackage{slashed}
\usepackage{siunitx}
\usepackage{amssymb}
\usepackage{graphicx}
\usepackage{epic}
\usepackage{eepic}
\usepackage{epsfig}
\usepackage{latexsym}
\usepackage[dvipsnames]{xcolor}
\usepackage[export]{adjustbox}
\usepackage{float}
\usepackage{multirow, makecell,changepage}
\usepackage{hyperref}
\usepackage{enumitem}
\hypersetup{colorlinks=true,citecolor=red,linkcolor=NavyBlue,urlcolor=NavyBlue}
\usepackage[caption=false]{subfig}
 
\usepackage{natbib}
\usepackage{relsize}
\usepackage{wrapfig}
\usepackage[left=1.65cm,right=1.65cm,top=1.9cm,bottom=1.9cm]{geometry}
\usepackage{slashed}

\begin{document}
\relscale{1.05}

\title{Vectorlike $\tau$ production through leptoquarks}

\author{Shruti Dubey}
\email{shruti22@iisertvm.ac.in}
\affiliation{Indian Institute of Science Education and Research Thiruvananthapuram, Vithura, Kerala, 695 551, India}

\author{Nilanjana Kumar}
\email{nilanjana.kumar@gmail.com}
\affiliation{Centre for Cosmology and Science Popularization, SGT University, Gurugram, Haryana, 122 505, India}

\author{Tanumoy Mandal}
\email{tanumoy@iisertvm.ac.in}
\affiliation{Indian Institute of Science Education and Research Thiruvananthapuram, Vithura, Kerala, 695 551, India}

\author{Subhadip Mitra}
\email{subhadip.mitra@iiit.ac.in}
\affiliation{Center for Computational Natural Science and Bioinformatics, 
International Institute of Information Technology, Hyderabad 500 032, India}

\affiliation{Centre for Quantum Science and Technology,
International Institute of Information Technology, Hyderabad 500 032, India.}

\author{Rachit Sharma}
\email{rachit21@iisertvm.ac.in}
\affiliation{Indian Institute of Science Education and Research Thiruvananthapuram, Vithura, Kerala, 695 551, India}

\begin{abstract}\noindent
Numerous phenomenological studies and collider searches have probed for the existence of new physics by looking for signatures of leptoquarks (LQs) or vectorlike leptons (VLLs). We consider a new possibility that can arise in theories with enhanced gauge symmetries: both particles are simultaneously present, and LQ-mediated processes enhance the VLL production at the LHC. We study the effect of non-standard interactions of LQs that contribute to novel production and decay signatures. We obtain the HL-LHC prospects of this framework in the mono-and di-lepton final states, and discuss other potentially relevant channels.
\end{abstract}

\maketitle 

\section{Introduction}\label{sec:intro}
\noindent
Searching for new uncoloured fermions at the Large Hadron Collider (LHC) poses significant challenges, primarily due to their small production cross sections. For instance, if the new fermion is a Standard Model (SM) gauge singlet such as the right-handed neutrino (RHN), its production at the LHC is suppressed by the small light-heavy neutrino mixing angle. If the fermion is instead uncoloured but has an electroweak charge like vector-like leptons (VLLs), it can be produced via electroweak interactions. However, the corresponding production rates will be much smaller than those achievable through strong interactions.  

An intriguing possibility arises in non-minimal extensions of the SM, where such uncoloured fermions coexist with other heavy bosonic states such as $W^\prime$, $Z^\prime$, or leptoquarks (LQs). If such a boson is produced copiously at the LHC, it could decay to these uncoloured fermions with a sizable branching fraction, making the boson production a production channel for the fermions. Studying the exotic decays of the heavy boson becomes interesting on its own, as the resulting collider signatures often contain final states and kinematic features not explored by the current LHC searches. Previously, we investigated the production of RHNs through the decays of new bosonic states~\cite{Choudhury:2020cpm,Deka:2021koh,ThomasArun:2021rwf,Arun:2022ecj,Bhaskar:2023xkm,Mandal:2023mck,Duraikandan:2024kcy}. In this paper, we extend this line of investigation by exploring the production of VLLs via LQs.

\begin{table*}[]
\caption{List of LQs and their representations under the SM gauge groups. The second column shows the representations under $SU(3)_c$, $SU(2)_L$, and $U(1)_Y$, respectively. The last column presents the renormalisable interactions of the LQs with the singlet VLL $\tau_2$.
\label{tab:1}}
\centering
{
\renewcommand\baselinestretch{2}\selectfont
\begin{tabular*}{\textwidth}{@{\extracolsep{\fill}} cccc}
\hline
\hline
LQ & ($SU(\mathbf{3})_{C}, SU(2)_{L}, U(1)_{Y}$) & Spin & Interaction Lagrangian \\
\hline \hline
$\mathbf{{S}_1}$ & $(\mathbf{\overline{3}},1,1/3)$& 0 & $   {y}_{1i}^{RR} \overline{u_{R}^{C}}^{i} {S}_{1} \tau_{2R} $ + h.c. \\
\hline
$\mathbf{\widetilde{S}_{1}}$ & $(\mathbf{\overline{3}},1,-2/3)$ & 0 & $  \widetilde{y}_{1i}^{{RR}} \overline{d_{R}^{C}}^{i}  \widetilde{S}_{1} \tau_{2R} $  + h.c. \\
\hline
$\mathbf{{R}_{2}}$ & $(\mathbf{3},2,1/6)$ & 0 & $ {y}_{2i}^{{LR}} \overline{Q}_{L}^{i,a} {R_{2}}^{a} \tau_{2R}$ + h.c.  \\
\hline
$\mathbf{{U}_1}$ & $(\mathbf{3},1,2/3)$& 1 & ${x}_{1i}^{RR} \overline{d}_{R}^{i} \gamma^{\mu} {U}_{1,\mu} \tau_{2R}$ + h.c. \\
\hline
$\mathbf{\widetilde{U}_{1}}$ & $(\mathbf{3},1,-1/3)$ & 1 & $  \widetilde{x}_{1i}^{{RR}} \overline{u}_{R}^{i} \gamma^{\mu} \widetilde{U}_{1,\mu} \tau_{2R}$  + h.c. \\
\hline
$\mathbf{{V}_{2}}$ & $(\mathbf{\overline{3}},2,-1/6) $& 1 & $   {x}_{2i}^{LR} \overline{Q_{L}^{C}}^{i,a} \gamma^{\mu} \varepsilon^{ab} {V}_{2,\mu}^{b}  \tau_{2R}$ + h.c. \\
\hline
\end{tabular*}
}
\end{table*}

VLLs are heavy partners of the SM leptons that belong to non-chiral representations of the SM gauge group, and hence, do not induce gauge anomalies. After electroweak symmetry breaking (EWSB), VLLs can mix with SM leptons. Depending on the generation of the SM lepton a VLL mixes with, we can assign a generation label to the VLL. Thus, a first-generation charged VLL mixes with an electron. Second-generation VLLs have been extensively studied in the literature, particularly in the context of the anomalous magnetic moment of the muon (see Refs.~\cite{Poh:2017tfo,Crivellin:2020ebi,Hamaguchi:2022byw}). In general, VLL phenomenology has been an active area of research~\cite{delAguila:2008pw,Martin:2009bg,FileviezPerez:2011pt,Joglekar:2012vc,Kearney:2012zi,Kumar:2015tna,Bhattacharya:2018fus,Chakraborty:2021tdo,Cherchiglia:2021syq,Bigaran:2023ris,Cingiloglu:2024vdh,Kumar:2025aek}. Mixing angles between VLLs and first- or second-generation SM leptons should be small, leading to long lifetimes and displaced-vertex signatures, as discussed in Refs.~\cite{Bernreuther:2023uxh,Bandyopadhyay:2023joz,Cao:2023smj}. In contrast, mixing with the $\tau$ lepton can be larger, resulting in prompt VLL decays. In this paper, we focus on third-generation charged VLLs, i.e. a heavy partner of $\tau$.

Current ATLAS searches have excluded vectorlike electron (muon) masses below $1220$~GeV ($1270$~GeV) and $320$~GeV ($400$ GeV) in the doublet and singlet scenarios, respectively~\cite{ATLAS:2024mrr}. For third-generation VLLs, the excluded limits from the LHC range from $100$–$1045$~GeV for doublets and $125$–$150$ GeV for singlets~\cite{CMS:2022nty,ATLAS:2023sbu}. In this work, we focus on weak-singlet third-generation VLLs, as the corresponding mass limits are relatively less stringent. We explore signatures of a non-minimal framework involving both VLLs and LQs. Being colour-charged, LQs can be produced copiously at the LHC via QCD-driven processes. They are well-motivated in a variety of Beyond the SM (BSM) frameworks, including the Pati–Salam model~\cite{Pati:1973uk,Pati:1974yy}, Grand Unified Theories~\cite{Georgi:1974sy,Fritzsch:1974nn}, composite models~\cite{Schrempp:1984nj}, colored Zee-Babu models~\cite{Kohda:2012sr}, technicolor models~\cite{Dimopoulos:1979es,Farhi:1980xs}, and supersymmetry with $R$-parity violation~\cite{Barbier:2004ez}. Many studies have investigated their collider phenomenology (see Refs.~\cite{Mandal:2015vfa,Mandal:2018kau,Aydemir:2019ynb,Chandak:2019iwj,Bhaskar:2020kdr,Bhaskar:2020gkk,Bhaskar:2021pml,Bhaskar:2021gsy,Bandyopadhyay:2021pld,Bhaskar:2022vgk,Aydemir:2022lrq,Bhaskar:2023ftn,Cheung:2023gwm,Bhaskar:2024swq,Bhaskar:2024snl,Bhaskar:2024wic,Das:2025osr} for a nonexhaustive list), and the current LHC data place their mass-limits roughly around $1.5$~TeV for sLQs and $2$~TeV for vLQs (see the summary plots from the ATLAS~\cite{ATL-PHYS-PUB-2024-012} and CMS~\cite{CMSPlot} collaborations for the current limits).

While most LQ studies focus on their direct decays to SM fermions (a lepton and a quark), we investigate non-standard decay channels where LQs dominantly decay to VLLs and SM quarks. Such scenarios naturally arise in some Pati–Salam-type unification models often referred to as the $4321$ models, where both VLLs and LQs coexist~\cite{DiLuzio:2017vat,Calibbi:2017qbu,DiLuzio:2018zxy}. Ref.~\cite{CMS:2022cpe} considered the $4321$ model as the benchmark where a VLL is produced via QED interactions, and the final state arises from an off-shell LQ, leading to a three-body VLL decay. In contrast, in our case, VLLs are produced from LQ decays and subsequently decay into SM particles, resulting in interesting collider signatures and kinematics. If the LQ-VLL-quark coupling is large, the production of VLL through LQ becomes important even at high LQ mass.  

If LQs predominantly decay into VLLs, existing collider constraints become largely inapplicable, leaving this region of parameter space unexplored. This opens a unique opportunity to search for novel LHC signatures. Production of VLLs through LQs becomes significant when both particles are kinematically accessible and the VLL is lighter than the LQ. Given the large LQ production cross section at the LHC, this mechanism can substantially enhance the VLL signal. The phenomenology in this channel depends on the VLL and LQ masses, the strength of the LQ-VLL-quark couplings, and the generation of the initial-state quarks.

We focus on scenarios where LQs couple to first-generation quarks and VLLs that mix with third-generation SM charged lepton, though the framework can be easily extended to other flavour structures. We include VLL production via pair production (PP), single production (SP), and indirect production (IP) of LQs~\cite{Bhaskar:2023ftn,Das:2025osr}. The produced VLLs subsequently decay into various SM states.
Focusing on mono-lepton and di-lepton final states, we derive exclusion and discovery contours for the relevant parameter space. We also identify several other promising final states for future investigation.

The remainder of this paper is organised as follows.  
In Sec.~\ref{Model}, we present our model with LQs and VLL and their relevant interactions.  
Sec.~\ref{pheno} describes different production mechanisms of LQs and VLLs.    
The signal and background selection criteria are detailed in Sec.~\ref{signal}, followed by the HL-LHC prospects discussed in Sec.~\ref{hllhc}.  
Finally, we summarise our findings and present our conclusions in Sec.~\ref{summary}.

\begin{figure*}
    \centering
    \captionsetup[subfigure]{labelformat=empty}
    \subfloat[\quad\quad\quad(a)]{\includegraphics[width=0.44\textwidth]{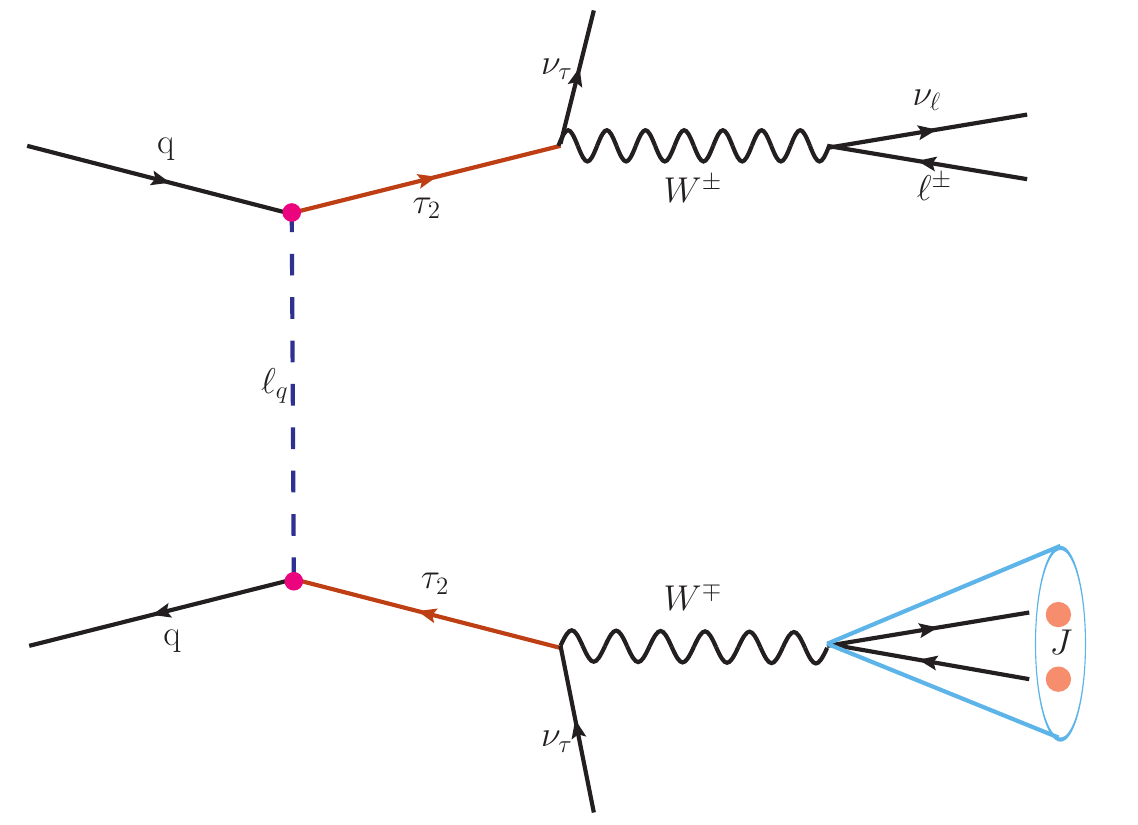}\label{feyn_mono}}\hspace{1cm}
    \subfloat[\quad\quad\quad(b)]{\includegraphics[width=0.4\textwidth]{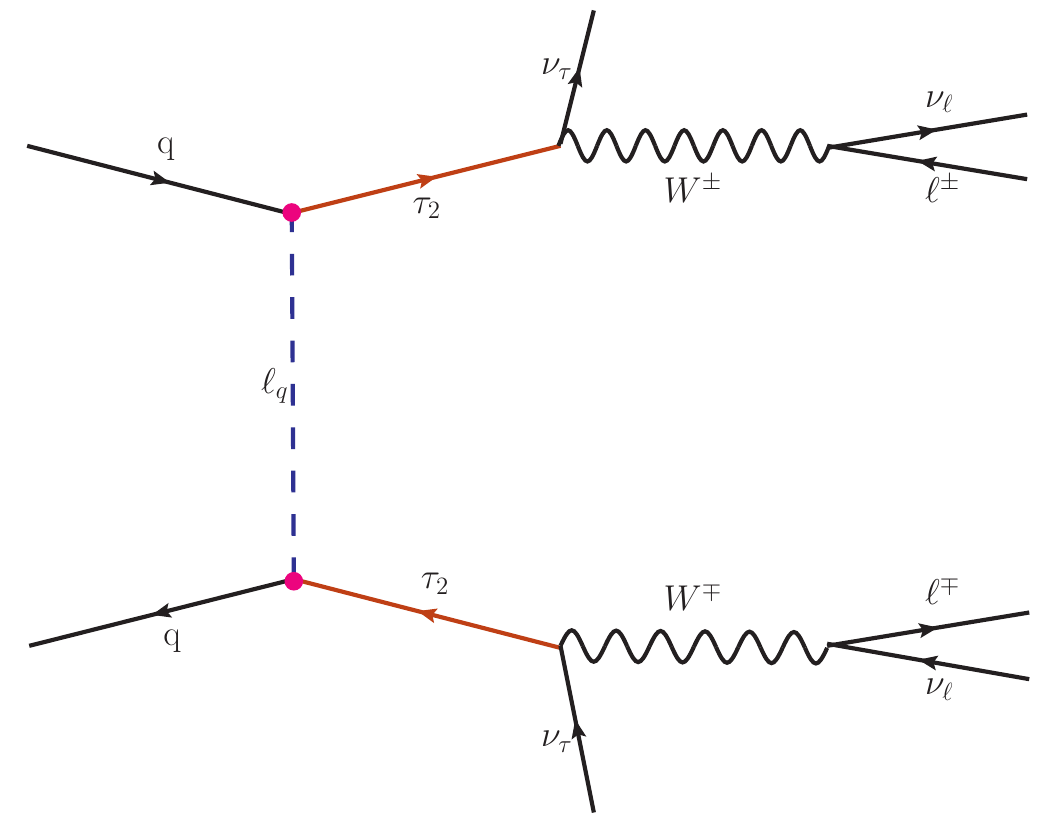}\label{feyn_di}}   
    \caption{Mono- and di lepton final states from $\tau_2$ pair production through a $t$-channel LQ exchange at the LHC.
 \label{fig:Feyn}}
\end{figure*}

\section{The Model} 
\label{Model}
\noindent
The interaction Lagrangian involving a singlet VLL before EWSB can be written as
\begin{align}
\label{eq:Lagsingvlqgen}
\mathcal{L} \supset &- \Big\{\widetilde{\lambda}_{\tau}(\overline{L}_{3L}H)\Lambda_R 
+ \omega_T^{\prime}(\overline{L}_{3L} H)T_{R}
+ \widetilde{\omega}_T m_T\, \overline{T}_{L} \Lambda_R \nonumber \\
& + M_{T} \overline{T}_{L} T_{R} 
+ \text{h.c.}\Big\},
\end{align}
where $T$ denotes the third-generation singlet VLL, $\Lambda_R$ is the third-generation right-handed SM  lepton, and $L_{3L}$ and $H$ are the third-generation SM lepton and Higgs doublets, respectively. The parameters $\widetilde{\lambda}_{\tau}$, $\omega_T^\prime$, and $\widetilde{\omega}_T$ are dimensionless couplings. The mass parameter $M_T$ corresponds to the VLL mass scale, which is not generated via the Higgs mechanism. An additional scale $m_T$ appears in the off-diagonal mixing term $\widetilde{\omega}_Tm_T\overline{T}{L} \Lambda_R$. This mixing term is gauge invariant and connects two particles having the same quantum numbers. This term may originate from high-scale symmetry breaking. However, if one does not track its UV origin, it becomes redundant and can be rotated away through an orthogonal transformation in the ${\Lambda_{R}, T_{R}}$ basis:
\begin{align}
\label{eq:transformation1}
T_{R} = 
 \tau_{R}^{\prime} \cos\theta
-  \tau_R \sin\theta, \quad
\Lambda_{R} = 
 \tau_{R}^{\prime} \sin\theta
+  \tau_R \cos\theta,
\end{align}
with the mixing angle $\theta = \tan^{-1}\left(\widetilde{\omega}_T m_T/M_T\right)$.
Just for notational convenience, we rename $T_L$ as $\tau^\prime_L$. The redefined Yukawa couplings are
\begin{align}
\lambda_\tau = \widetilde{\lambda}_\tau \cos\theta - \omega_T^{\prime} \sin\theta, \quad
\omega_T = \omega_T^{\prime}\cos\theta + \widetilde{\lambda}_\tau \sin\theta,
\end{align}
while the effective VLL mass scale becomes
\begin{equation}
M_{\tau^{\prime}} = \widetilde{\omega}_T m_T \sin\theta + M_T \cos\theta.
\end{equation}

\noindent
In this new basis, the Lagrangian becomes,
\begin{align}
\label{eq:Lagpheno}
\mathcal{L} \supset - \Big\{\, 
\lambda_\tau(\overline{L}_{3L} H) \tau_R 
+ \omega_T(\overline{L}_{3L} H)\tau_{R}^{\prime}
+ M_{\tau^\prime} \overline{\tau}_{L}^{\prime} \tau_{R}^{\prime} 
+ \text{h.c.} \,\Big\}.
\end{align}
After EWSB, the mass Lagrangian takes the form:
\begin{align}
\mathcal{L} \supset -
\begin{pmatrix} \overline{\tau}_L & \overline{\tau}_{L}^{\prime} \end{pmatrix}
\begin{pmatrix}
\lambda_\tau \, \dfrac{v}{\sqrt{2}} & \omega_T \, \dfrac{v}{\sqrt{2}} \\
0 & M_{\tau^\prime}
\end{pmatrix}
\begin{pmatrix}
\tau_R \\ \tau_{R}^{\prime}
\end{pmatrix}
+ \text{h.c.},
\label{eq:massmats}
\end{align}
where $v$ is the vacuum expectation value (VEV) of the Higgs field. Similar parameterisation in the context of vector-like quarks is already mentioned in Ref.~\cite{Bhardwaj:2022nko}. By means of a bi-orthogonal transformation, one can go from the interaction basis $\tau,\tau^\prime$ to the physical basis $\tau_1,\tau_2$. The interactions of $\tau_2$ with the SM gauge bosons via mixing are presented in Appendix~\ref{app:tau2_gauge}.


We consider all sLQs and vLQs that can couple to a singlet VLLs. We assume that the couplings of LQs to SM quarks and leptons are negligible compared to their couplings to VLLs and quarks. Table~\ref{tab:1} summarises the relevant LQ representations, including their quantum numbers under the SM gauge group and their renormalisable interactions with the third-generation singlet VLL, denoted by $\tau_{2}$ (mass eigenstate).

We denote the coupling matrices of sLQs and vLQs with fermion by $y$ and $x$, respectively. The first subscript of these couplings indicates the $SU(2)_L$ representation of the LQ, while the superscripts represent the chiralities of the involved fermions, where the first index corresponds to the quark and the second to the lepton.
For simplicity, we take both the Cabibbo--Kobayashi--Maskawa (CKM) and the Pontecorvo--Maki--Nakagawa--Sakata (PMNS) mixing matrices to be approximately the identity. This approximation is justified, as the processes initiated by the first-generation quarks—unlike those involving second-generation ones~\cite{Bhaskar:2023xkm}—are not significantly affected by small off-diagonal elements in the CKM matrix. Moreover, the flavour of missing neutrinos cannot be distinguished in LHC detectors.

vLQs can also possess additional interactions with gluons allowed by gauge invariance. These interactions can modify their production cross sections at hadron colliders~\cite{Blumlein:1994tu,Blumlein:1996qp}. The relevant interaction term in the Lagrangian is given by:
\begin{align}
\mathcal{L} \supset 
 - i g_{s} (1 - \kappa) \, \chi^{\dagger}_{\mu} T^{a} \chi_{\nu} \, G^{a\, \mu\nu},
\end{align}
where $\chi_{\mu}$ denotes a generic vLQ, $G^{a\, \mu\nu}$ is the gluon field strength tensor, $T^a$ are the $SU(3)_c$ generators, and $\kappa$ is a dimensionless parameter governing the strength of the vLQ-gluon interaction. Throughout this work, we set $\kappa = 1$ in all our results. The dominant contribution to the signal arises from the indirect production (IP) channel, which is largely insensitive to the value of $\kappa$. While varying $\kappa$ can slightly shift the coupling-independent mass bounds, it does not significantly affect the overall results.

\section{LHC Phenomenology} 
\label{pheno}
\noindent
 We generate the Universal FeynRules Output (UFO) files using \textsc{FeynRules}~\cite{Alloul:2013bka}, incorporating all the BSM particles along with their relevant interactions. The UFO model is then imported into \textsc{MadGraph5}~\cite{Alwall:2014hca} to simulate both signal and background events at leading order (LO). For event generation, we use the \textsc{NNPDF23LO1}~\cite{NNPDF:2021uiq} parton distribution functions with the default dynamical scale choice in \textsc{MadGraph5}. To account for higher-order corrections, we apply a $K_{QCD}$ factor of $1.58$~\cite{Kramer:2004df,Mandal:2015lca,Borschensky:2020hot,Borschensky:2021hbo,Borschensky:2022xsa} for sLQ PP. Similar higher-order effects are included for the various background processes, with their corresponding cross sections and orders listed in Table~\ref{tab:Backgrounds}. The generated events are subsequently processed through \textsc{Pythia8}~\cite{Bierlich:2022pfr} for parton showering and hadronisation, and through \textsc{Delphes3}~\cite{deFavereau:2013fsa} for detector simulation using the default CMS card. Jet clustering is performed with the \textsc{FastJet}~\cite{Cacciari:2011ma,Cacciari:2008gp} package. We use jets with two different radii, AK4 jets with radius $0.4$ (denoted by $j$) and fatjets with radius $0.8$ (denoted by $J$), in our analysis.

\begin{figure*}
    \centering
    \captionsetup[subfigure]{labelformat=empty}
    \subfloat[\quad\quad\quad(a)]{\includegraphics[width=0.42\textwidth]{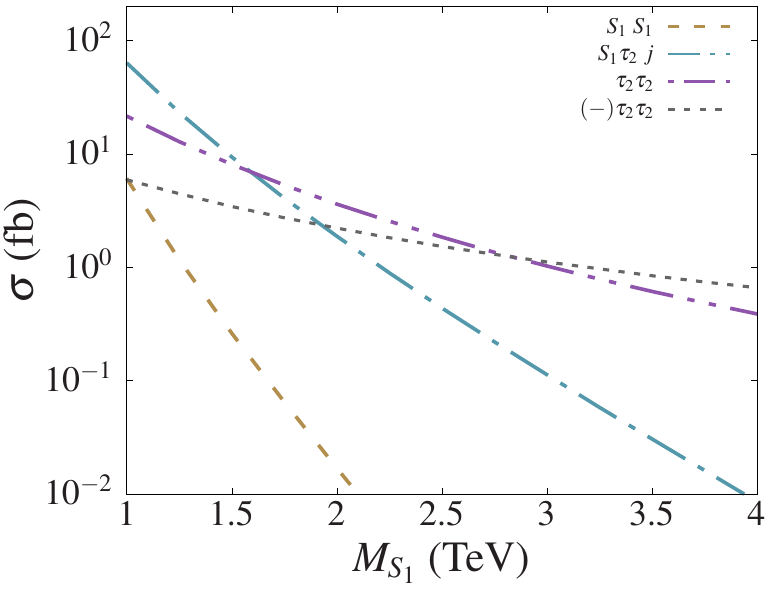}\label{S1_CS}}\hspace{1cm}
    \subfloat[\quad\quad\quad(b)]{\includegraphics[width=0.42\textwidth]{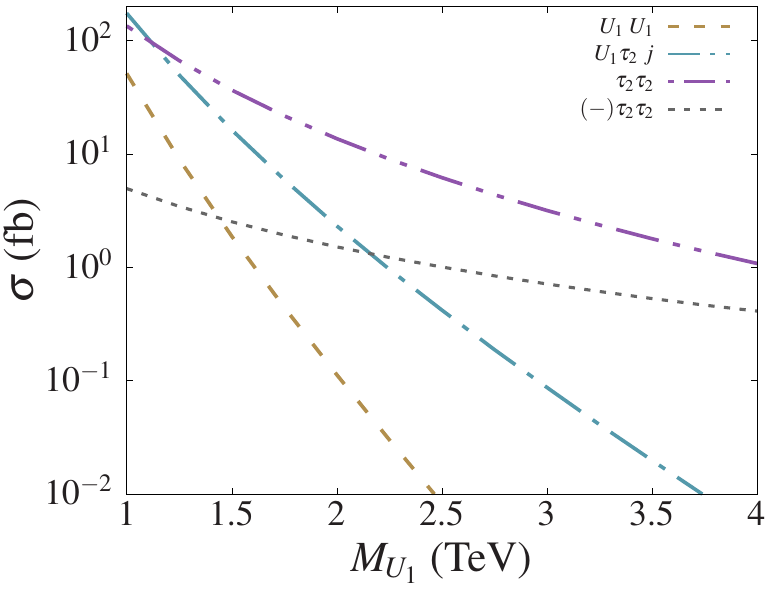}\label{U1_CS}}\\
    \subfloat[\quad\quad\quad(c)]{\includegraphics[width=0.42\textwidth]{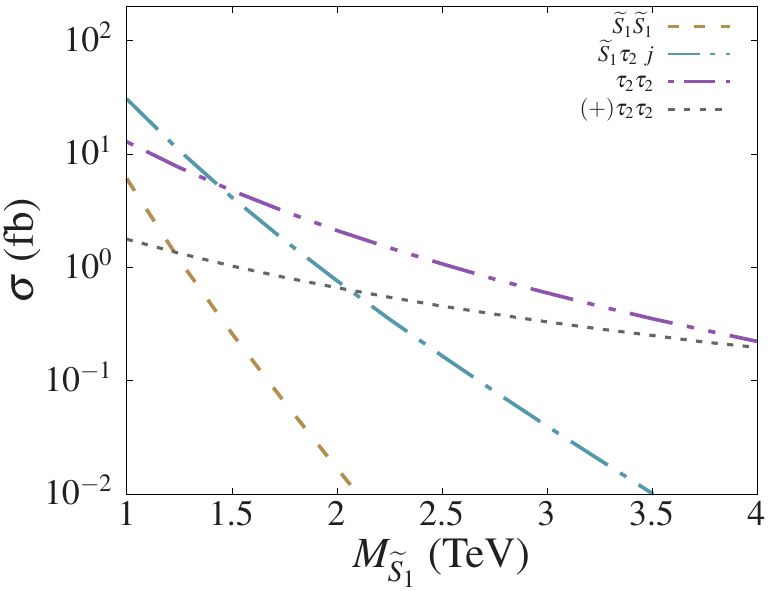}\label{S1t_CS}}\hspace{1cm}
    \subfloat[\quad\quad\quad(d)]{\includegraphics[width=0.42\textwidth]{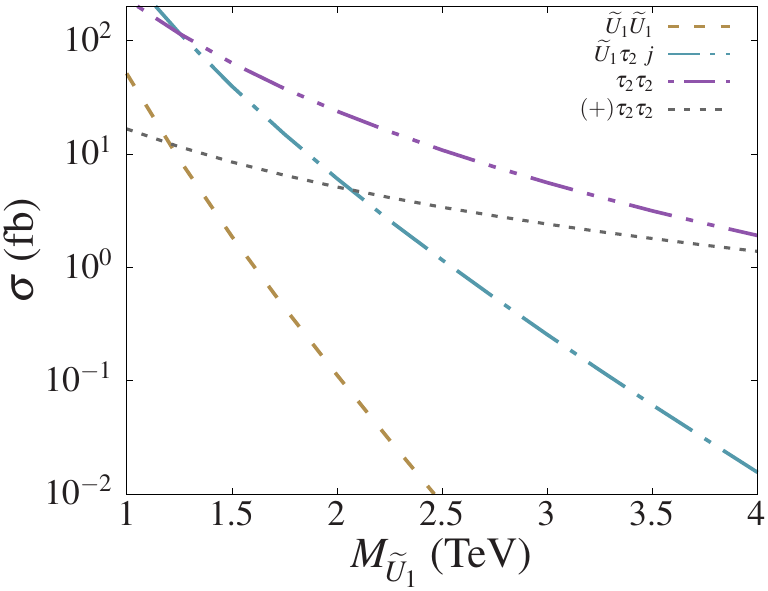}\label{U1t_CS}}   
    \caption{Cross sections for direct and indirect production modes of sLQs ($S_1$ and $\widetilde{S}_1$) and vLQs ($U_1$ and $\widetilde{U}_1$) at the $14$~TeV LHC. All plots are shown for $M_{\tau_2} = 500$~GeV with $x = y = 1$. The vLQ results assume $\kappa = 1$. Here, \((\pm) \tau_2\tau_2\) denotes the cross section for the II contribution with the corresponding sign.
 \label{MLQvsCS}}
\end{figure*}

\begin{table}[!b]
\caption{Higher-order cross sections of the dominant background processes, evaluated without including decays or selection cuts. The corresponding QCD orders are listed in the last column. These cross sections are used to compute the $K$-factors for incorporating higher-order corrections.\label{tab:Backgrounds}
}
\centering{\small\renewcommand\baselinestretch{1.35}\selectfont
\begin{tabular*}{\columnwidth}{l @{\extracolsep{\fill}} crc }
\hline
\multicolumn{2}{l}{Background } & $\sg$ & QCD\\ 
\multicolumn{2}{l}{processes}&(pb)&order\\\hline\hline
\multirow{2}{*}{$V +$ jets~ \cite{Catani:2009sm,Balossini:2009sa}  } & $Z +$ jets  &  $6.33 \times 10^4$& N$^2$LO \\ 
                & $W +$ jets  & $1.95 \times 10^5$& NLO \\ \hline
\multirow{3}{*}{$VV +$ jets~\cite{Campbell:2011bn}}   & $WW +$ jets  & $124.31$& NLO\\ 
                  & $WZ +$ jets  & $51.82$ & NLO\\ 
                   & $ZZ +$ jets  &  $17.72$ & NLO\\ \hline
\multirow{3}{*}{single $t$~\cite{Kidonakis:2015nna}}  & $tW$  &  $83.10$ & N$^2$LO \\ 
                   & $tb$  & $248.00$ & N$^2$LO\\ 
                   & $tj$  &  $12.35$ & N$^2$LO\\  \hline
$tt$~\cite{Muselli:2015kba}  & $tt +$ jets  & $988.57$ & N$^3$LO\\ \hline
\multirow{2}{*}{$ttV$~\cite{Kulesza:2018tqz}} & $ttZ$  &  $1.05$ &NLO+N$^2$LL \\
                   & $ttW$  & $0.65$& NLO+N$^2$LL \\ \hline
\end{tabular*}}
\end{table}

\subsection{Production at the LHC} 
\label{analysis}
\noindent
LQs can be produced at colliders via both resonant and non-resonant mechanisms. In the resonant case, they are either produced in pairs—referred to as pair production (PP)—or singly in association with a SM particle, known as single production (SP). Non-resonant production arises through $t$-channel LQ exchange between initial-state quarks, resulting in a pair of $\tau_2$ leptons; this is commonly referred to as indirect production (IP). Additionally, $\tau_2$ pairs can also be produced through standard QED processes via $s$-channel $Z/\gamma^*$ exchange. The IP and QED contributions can interfere, leading to an indirect interference (II) term, which can be either constructive or destructive depending on the LQ species. We emphasize that, in this work, II denotes the interference between LQ-mediated and QED-mediated $\tau_2$ pair production diagrams, whereas the II in Refs.~\cite{Bhaskar:2023ftn,Das:2025osr} refers to interference between LQ-mediated dilepton production and the SM dilepton background.

In the high-mass regime, non-resonant LQ production channels (IP and II) dominate over resonant channels (PP and SP), provided the new physics coupling is sufficiently large. The cross sections for the different production mechanisms as functions of the LQ mass are shown in Fig.~\ref{MLQvsCS}. PP is largely insensitive to the coupling, whereas SP and II scale with the square of the coupling. In contrast, the IP contribution scales with the fourth power of the coupling, making it particularly sensitive to large coupling values.

As mentioned earlier, we restrict our analysis to a third-generation singlet VLL, denoted by $\tau_2$. Among various LQs discussed in the literature, only $S_1$, $\widetilde{S}_1$, $R_2$, $U_1$, $\widetilde{U}_1$, and $V_2$ can couple to the singlet $\tau_2$. We denote a generic LQ state by $\ell_q$. Both resonant and non-resonant LQ production channels can give rise to a pair of $\tau_2$ leptons, which subsequently decay into a variety of observable final states at the LHC. We assume a mass hierarchy $M_{\ell_q} > M_{\tau_2}$, ensuring that the LQ decays to a $\tau_2$, and further assume the corresponding branching ratio to be nearly $100$\%. Among the several interesting final states, in this study, we focus on final states with one or two charged leptons. We consider both first- and second-generation leptons in the final states. Below, we discuss some interesting final states that can arise in our scenario.

\begin{table*}
\caption{Selection cuts applied on the monolepton and dilepton final states. \label{tab:Cuts}}
\centering
\renewcommand\baselinestretch{1.5}\selectfont
\begin{tabular*}{\textwidth}{@{\extracolsep{\fill}} lll}
\hline
\multicolumn{1}{r}{\multirow{1.5}{*}{Selection Cuts}} & \multicolumn{2}{c}{Channels}                   \\ \cline{2-3} 
\multicolumn{1}{r}{}                                & Monolepton         & Dilepton              \\ \hline \hline
C1                                                 & \begin{tabular}[c]{@{}l@{}}$p_{T}(\ell) > 60$ GeV, \\ No $b$-tagged jet \end{tabular}   & \begin{tabular}[c]{@{}l@{}} No $b$-tagged jet \end{tabular} \\ \hline
C2   & \begin{tabular}[c]{@{}l@{}}$M(J_1) > 40$ GeV,   \\  $\tau_{21}(J_1) < 0.5$ \end{tabular} & \begin{tabular}[c]{@{}l@{}}$p_{T}(J_1) > 200$ GeV \end{tabular}  \\  \hline
C3                                                  &  $\Delta \Phi (\slashed{E}_T,\ell) > 0.6$                 & \begin{tabular}[c]{@{}l@{}}$M(\ell_1,\ell_2) > 350$ GeV  \end{tabular}                  \\ \hline
C4                                                & \begin{tabular}[c]{@{}l@{}}$\slashed{E}_T >  300$ GeV\end{tabular} & $\slashed{E}_T >  450$ GeV\\ \hline
\end{tabular*}
\end{table*}

\begin{figure*}[]
    \centering
    \captionsetup[subfigure]{labelformat=empty}
    \subfloat[(a)]{\includegraphics[width=0.25\textwidth]{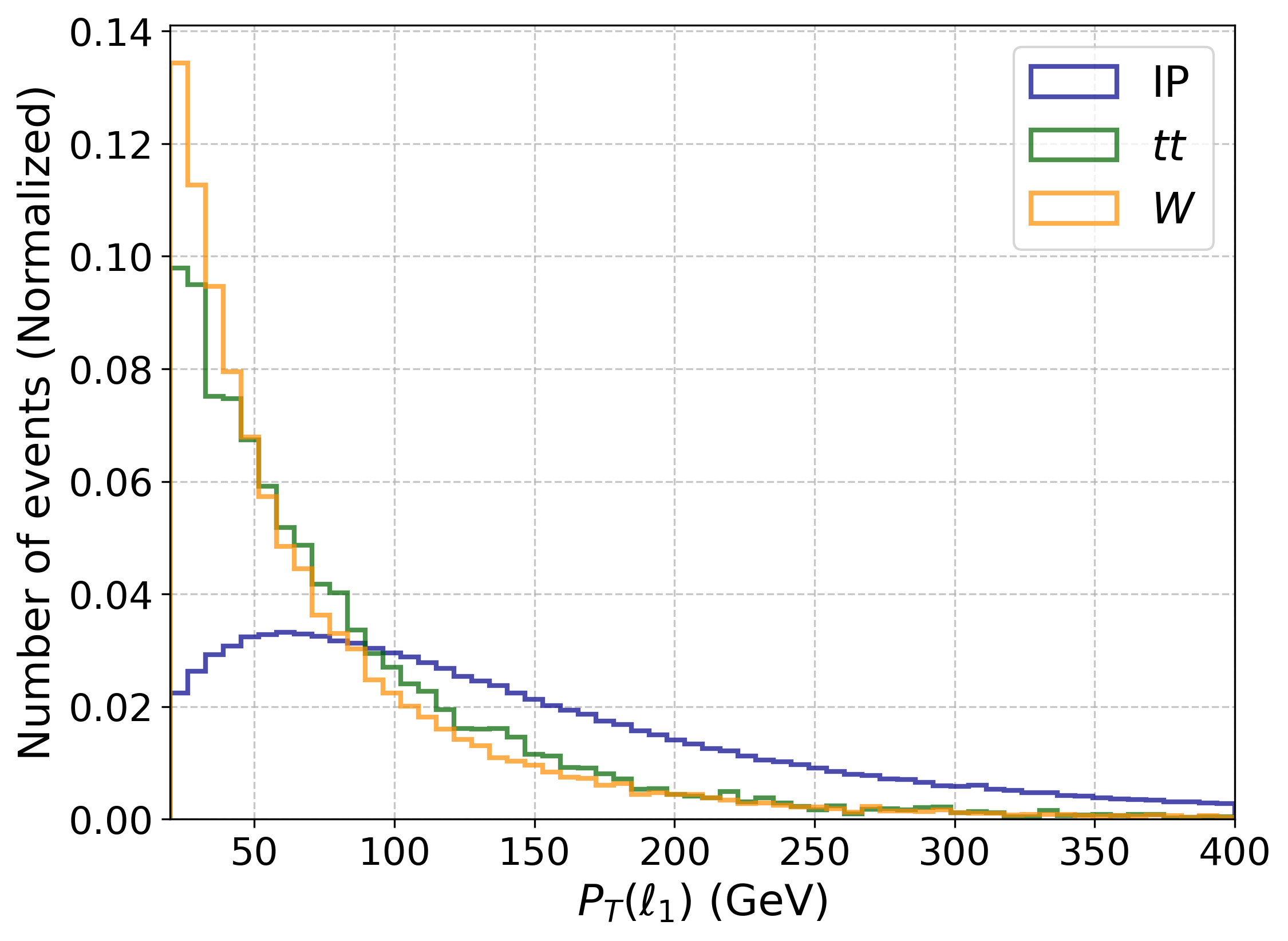}\label{pt_l_mono}}\hfill
    \subfloat[(b)]{\includegraphics[width=0.25\textwidth]{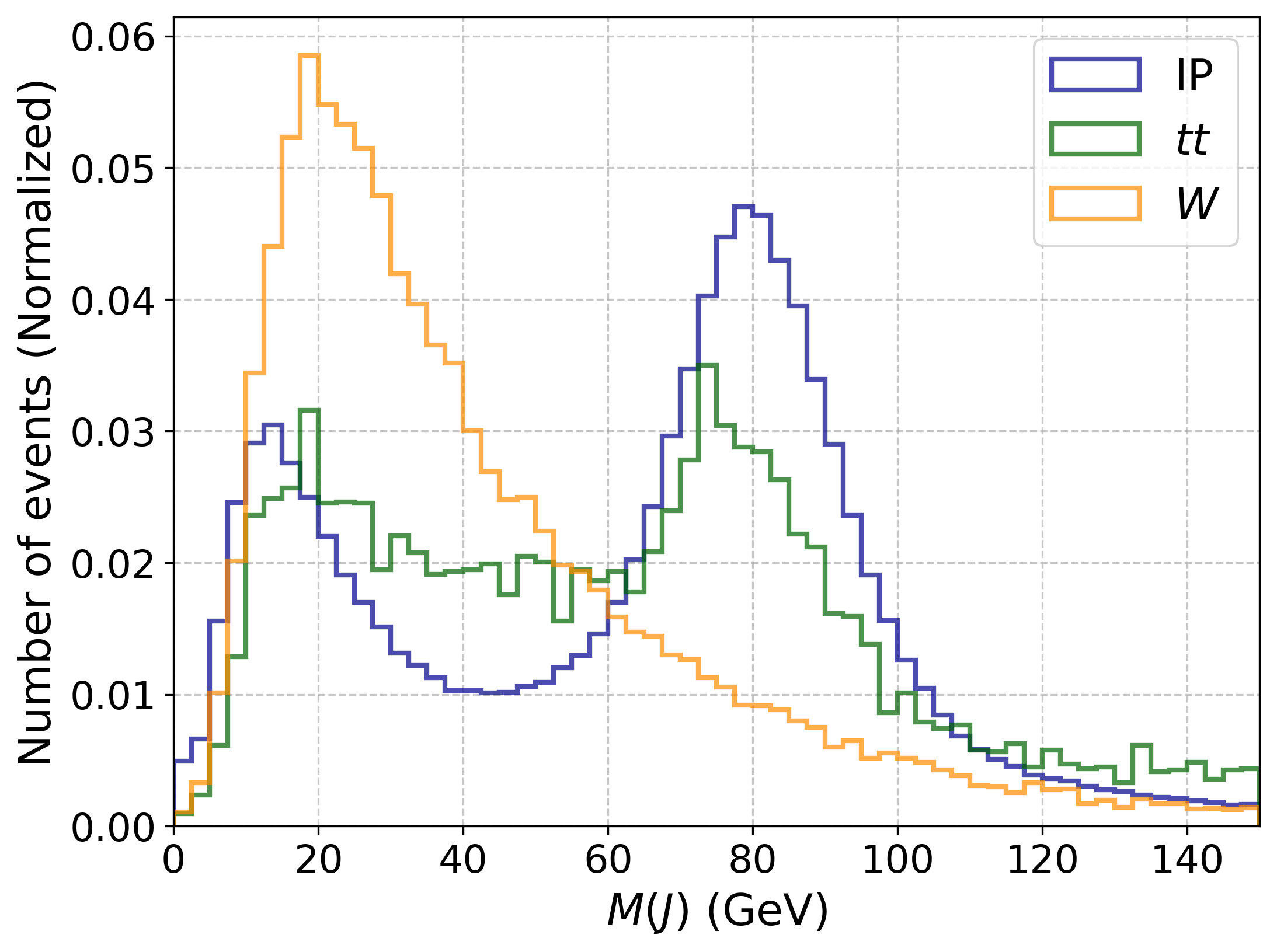}\label{pt_fj_mono}}\hfill
    \subfloat[(c)]{\includegraphics[width=0.25\textwidth]{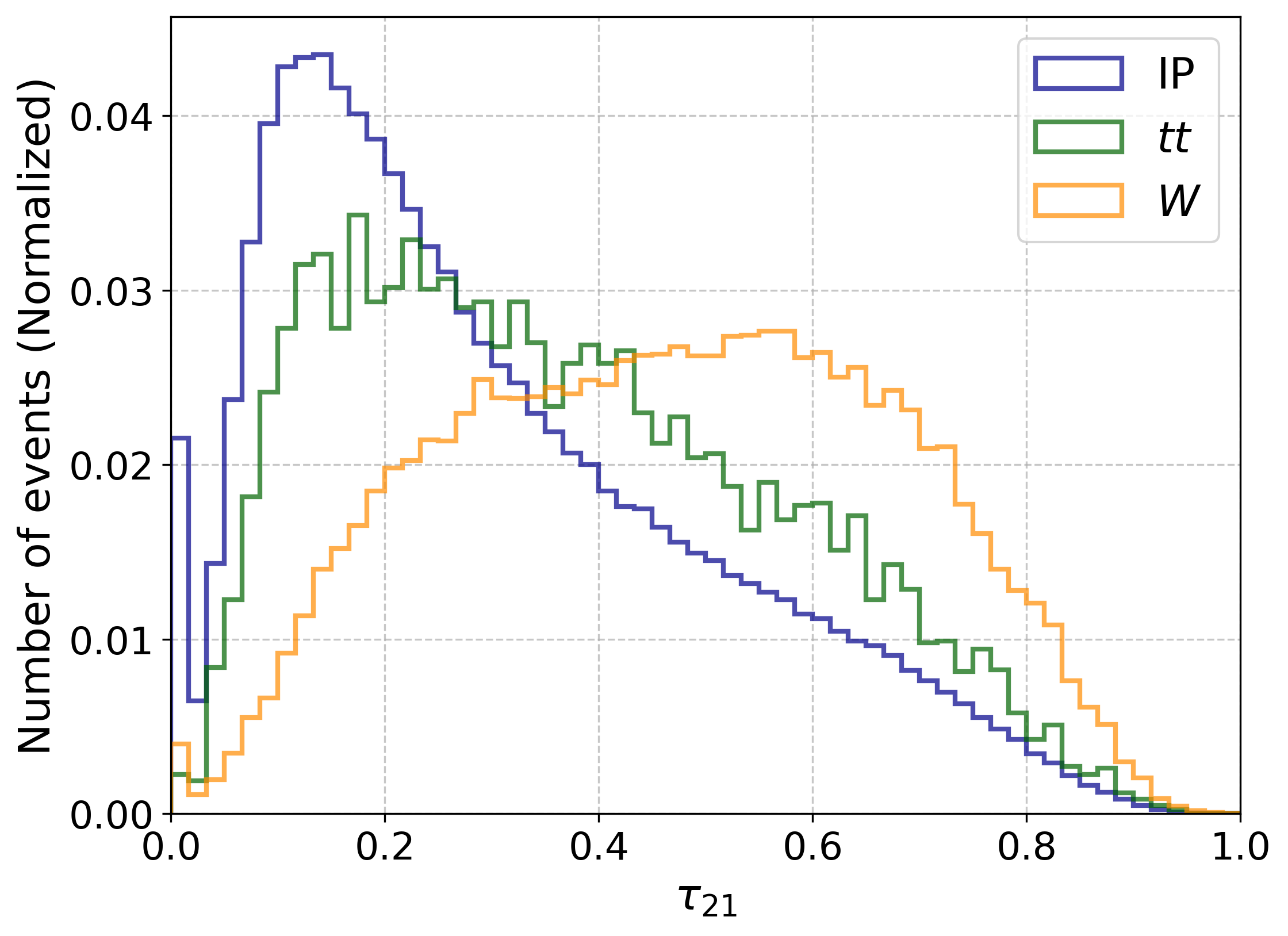}\label{tau21_mono}}\hfill
    \subfloat[(d)]{\includegraphics[width=0.25\textwidth]{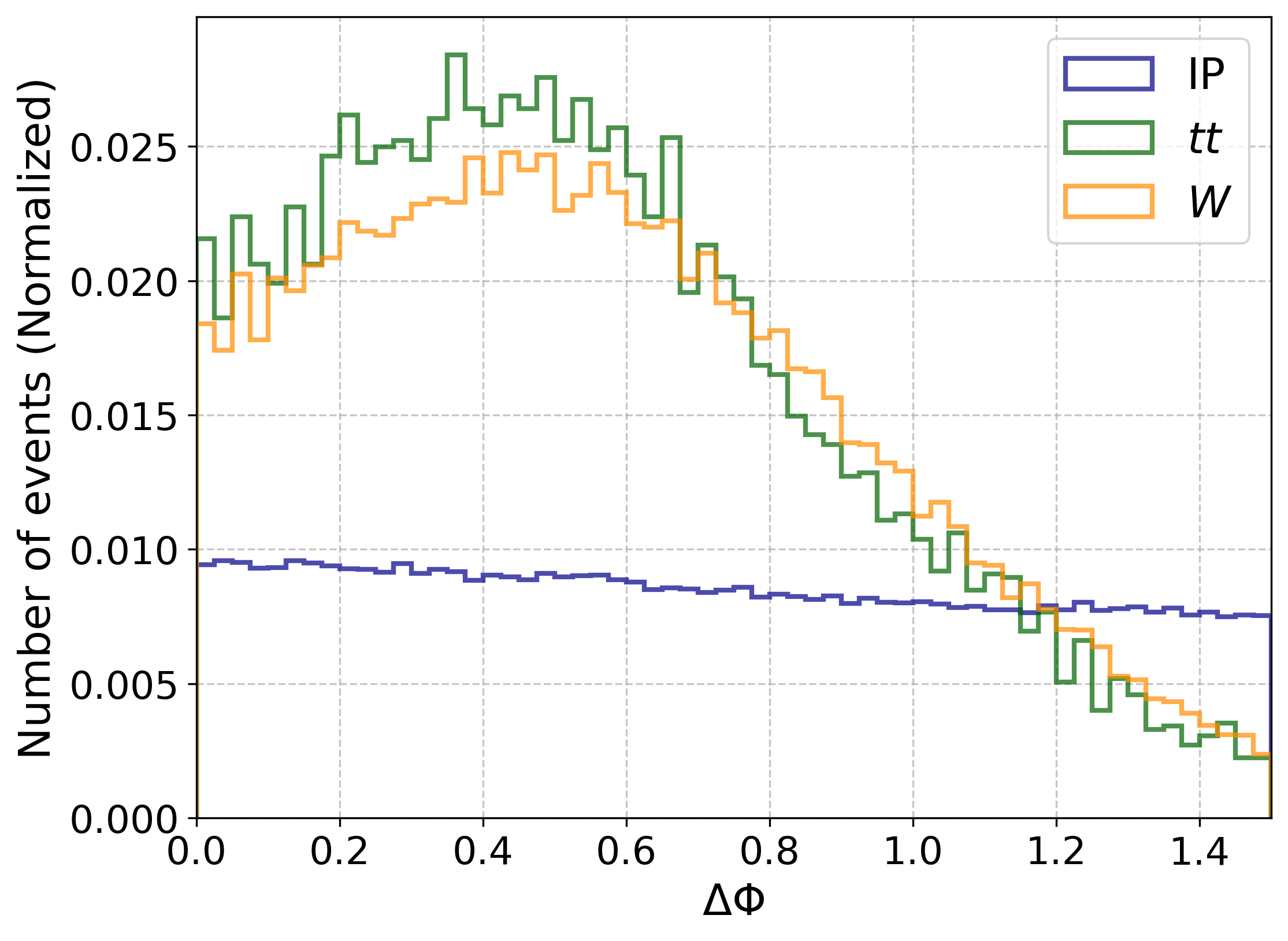}\label{delphi_mono}}\\
    \subfloat[(e)]{\includegraphics[width=0.25\textwidth]{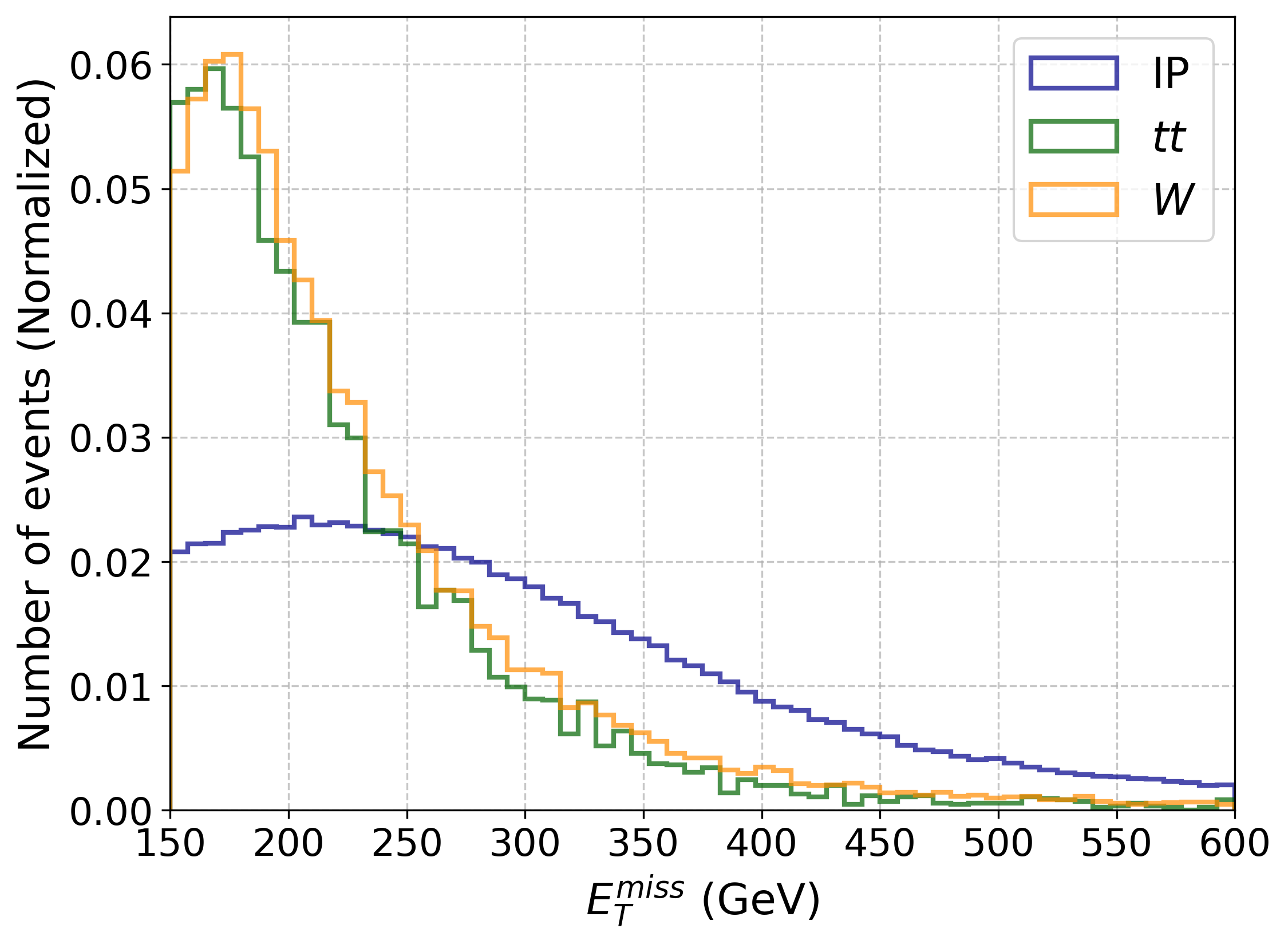}\label{met_mono}}\hfill
    \subfloat[(f)]{\includegraphics[width=0.25\textwidth]{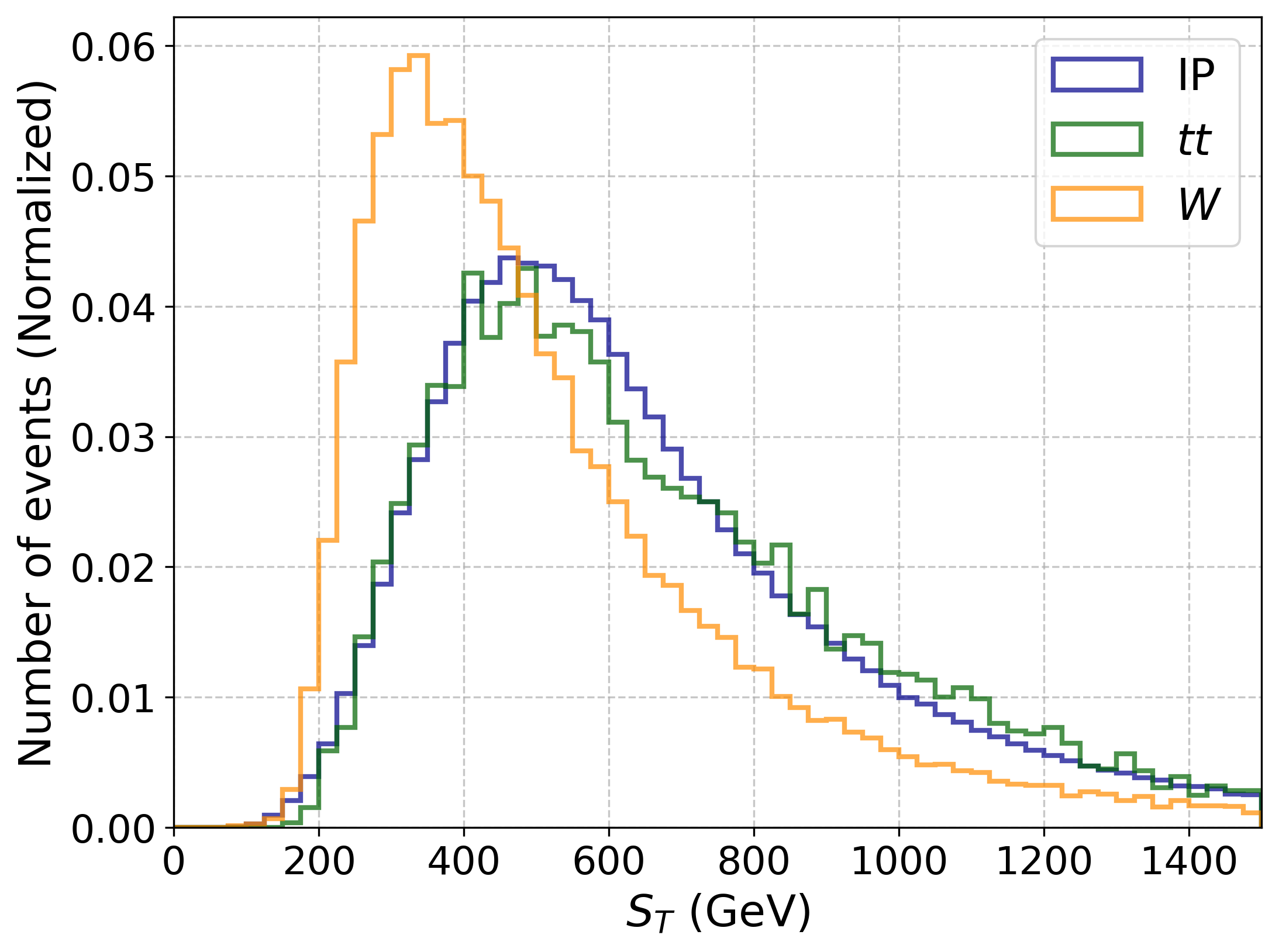}\label{ST_mono}}\hfill
    \subfloat[(g)]{\includegraphics[width=0.25\textwidth]{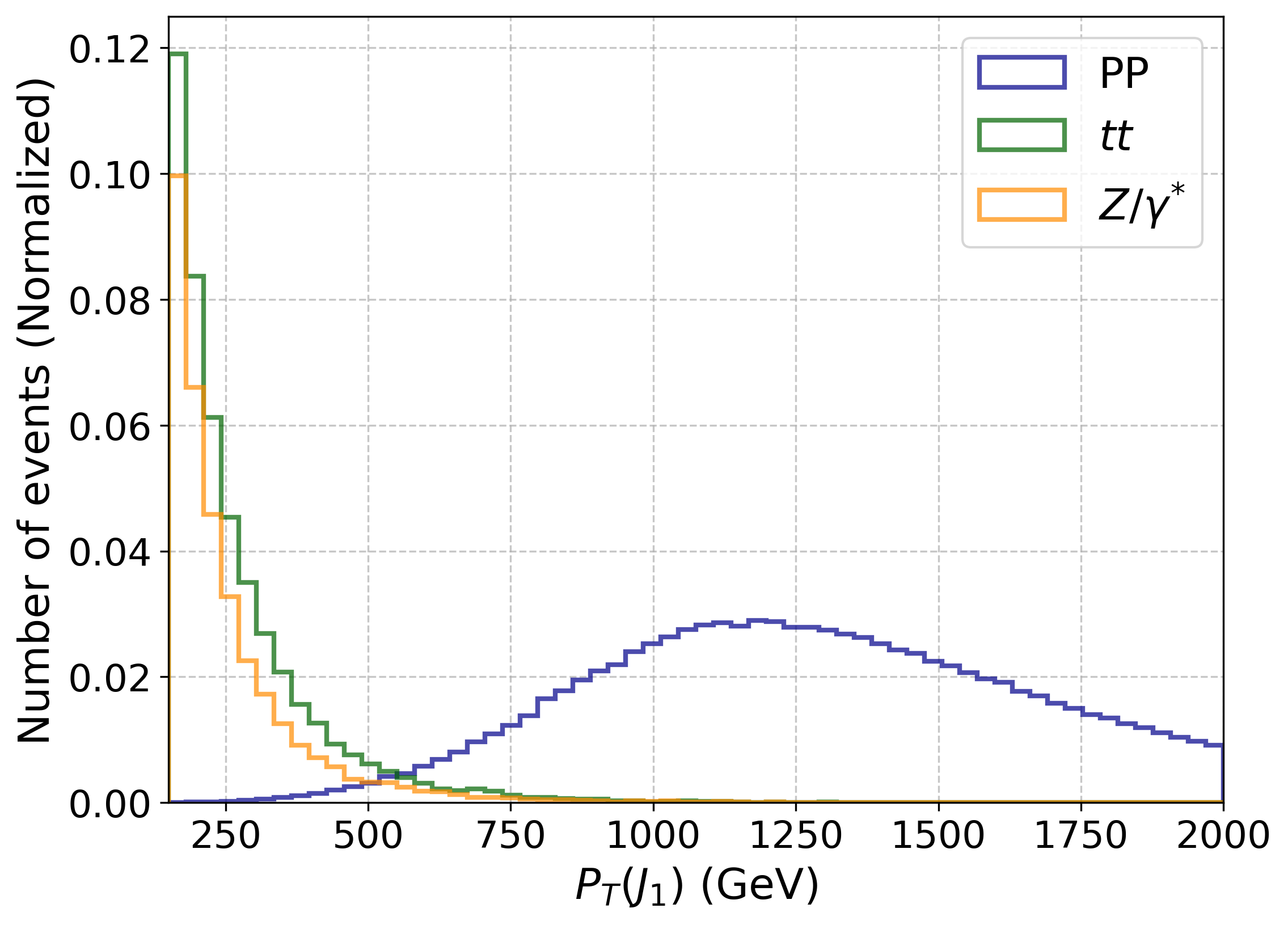}\label{pt_fj_dilep}}\hfill
    \subfloat[(h)]{\includegraphics[width=0.25\textwidth]{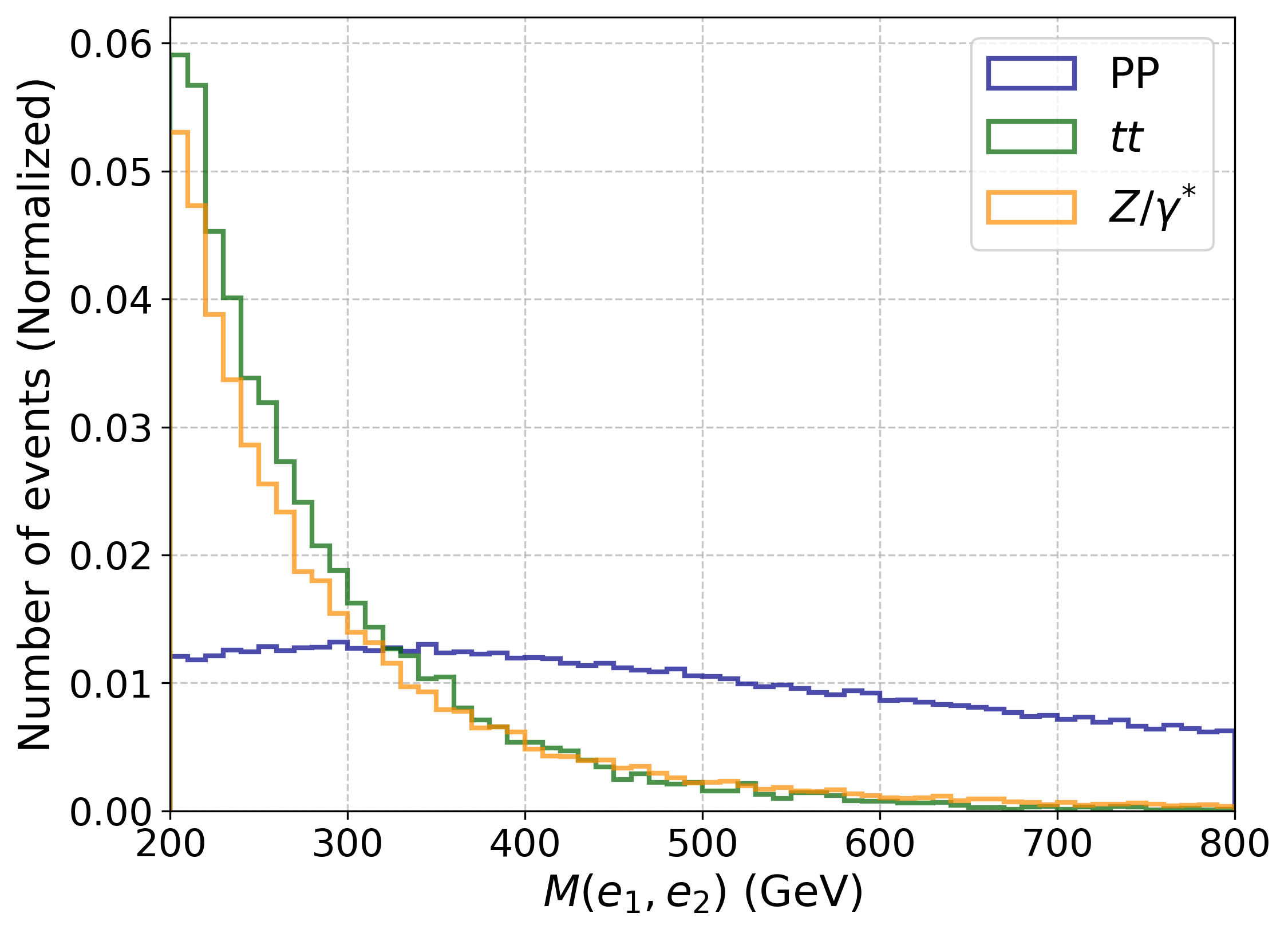}\label{IM_ll_dilep}}\\
    \subfloat[(i)]{\includegraphics[width=0.25\textwidth]{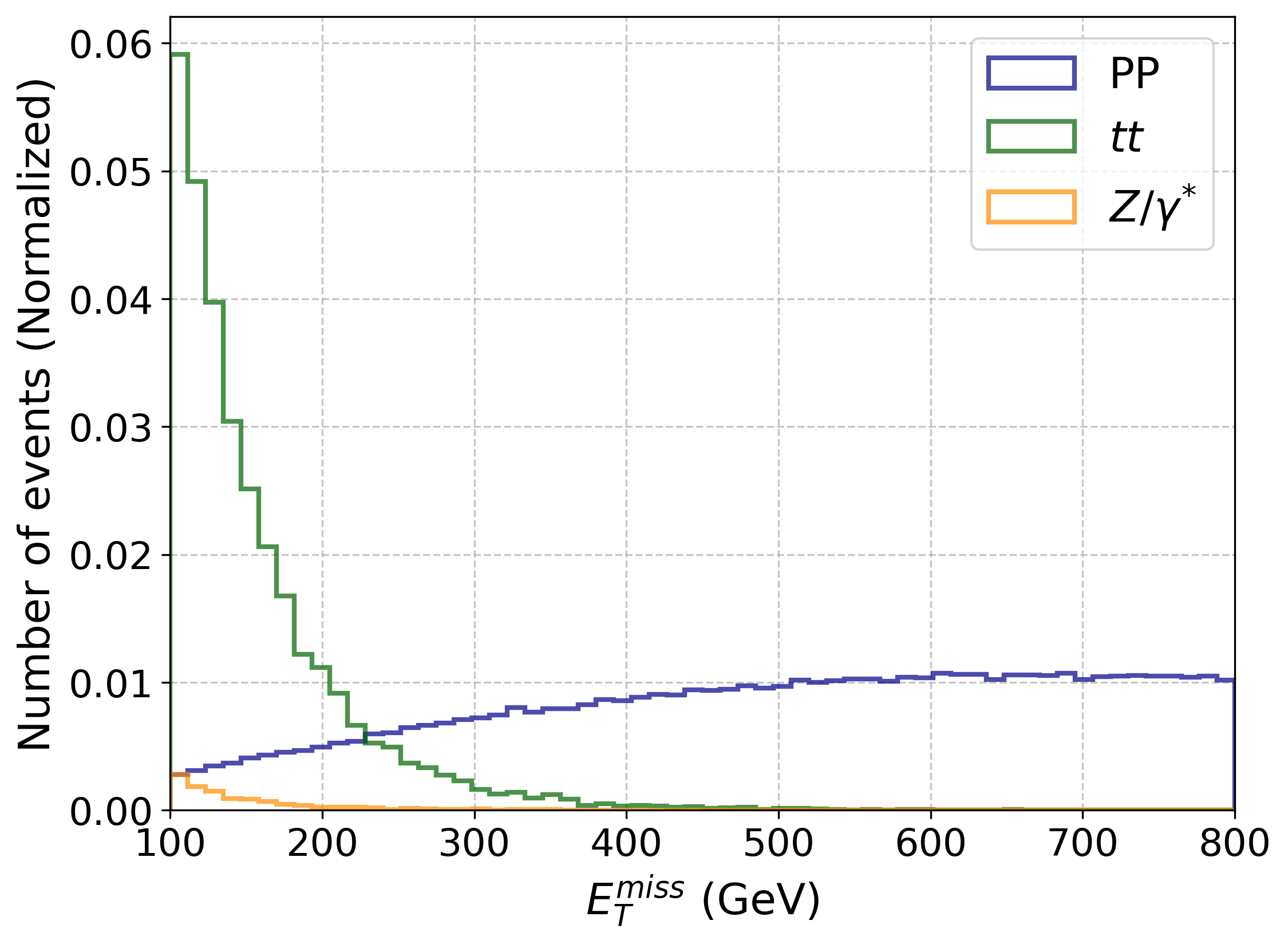}\label{met_dilep}}\hfill
    \subfloat[(j)]{\includegraphics[width=0.25\textwidth]{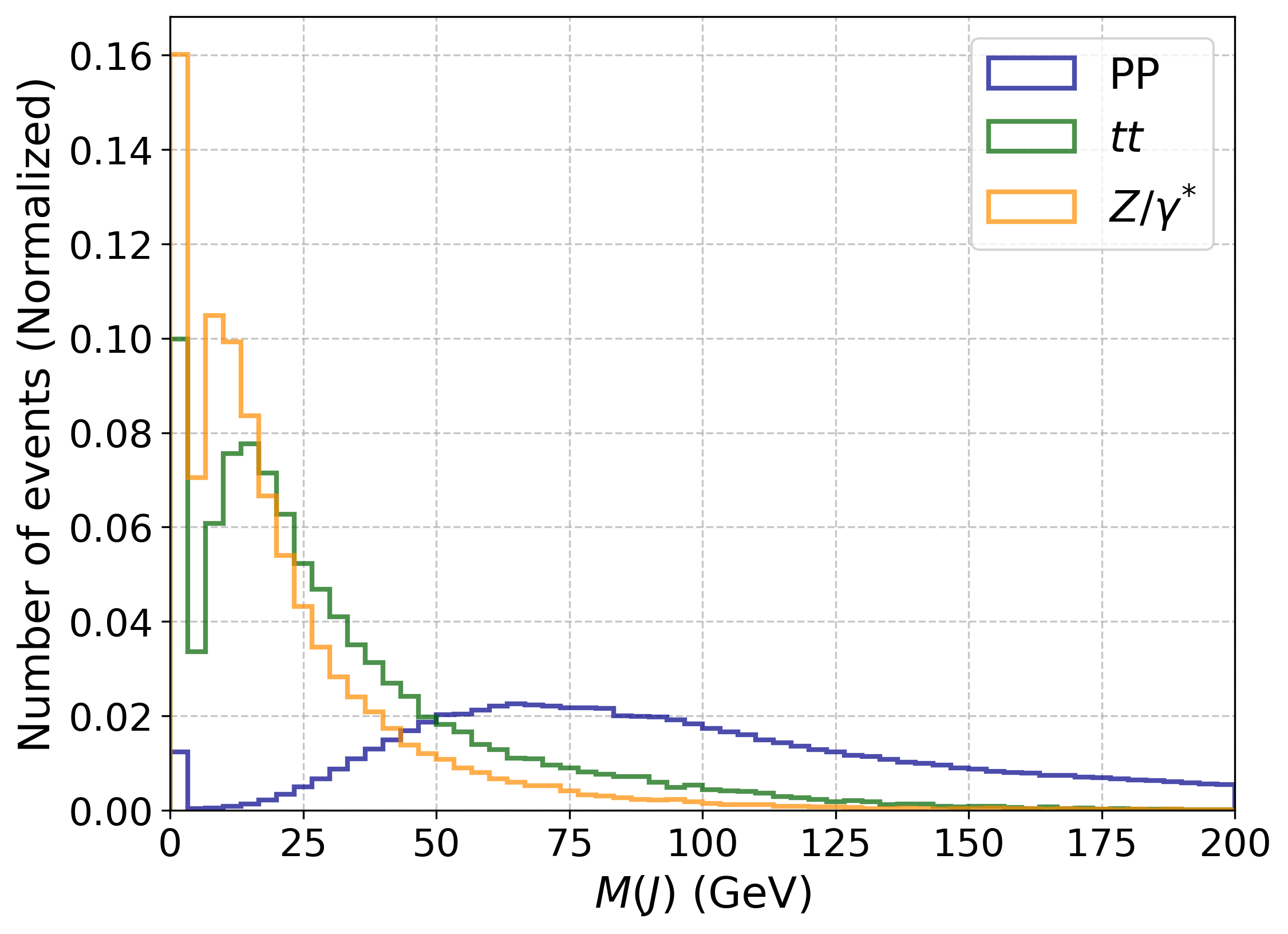}\label{M_fj_dilep}}\hfill
    \subfloat[(k)]{\includegraphics[width=0.25\textwidth]{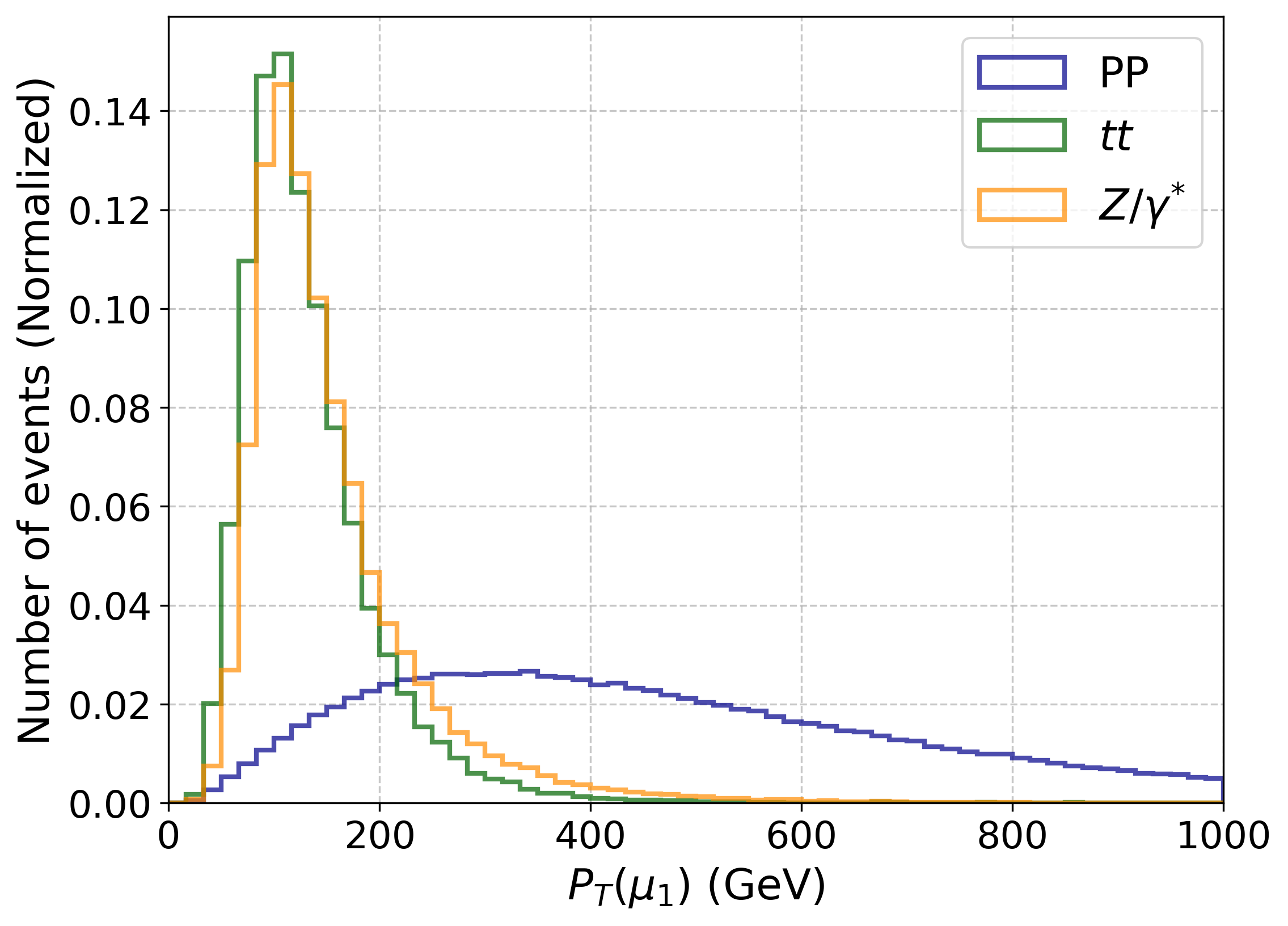}\label{pt_lp1_dilep}}\hfill
    \subfloat[(l)]{\includegraphics[width=0.25\textwidth]{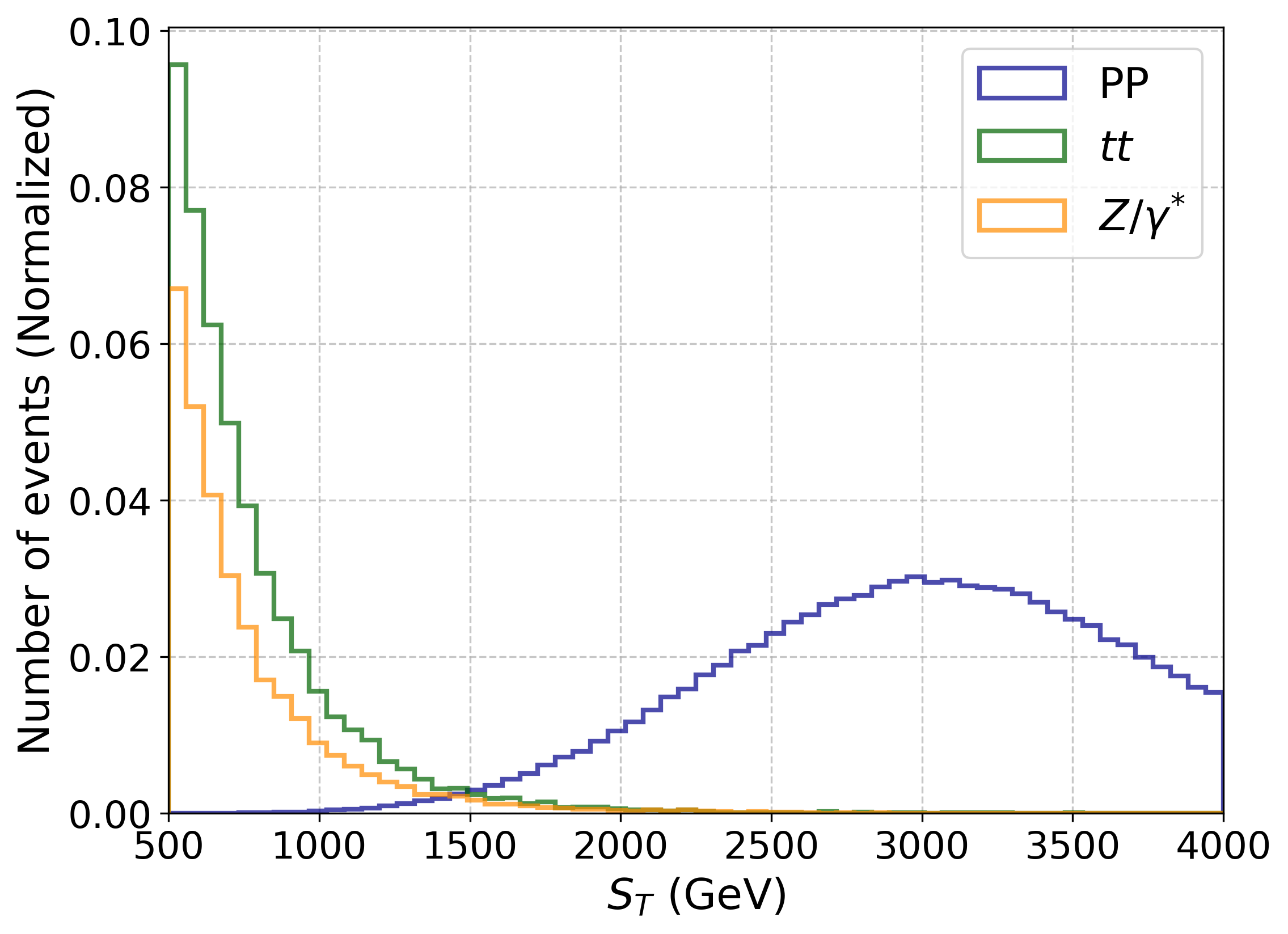}\label{ST_dilep}}
\caption{Normalised distributions of kinematic variables for the signal and two dominant backgrounds in monolepton [(a)--(f)] and dilepton [(g)--(l)] final states. The signal benchmark is ($M_{U_{1}}, M_{\tau_{2}}$) = (2.0~TeV, 0.5~TeV). For monolepton, the dominant backgrounds are $W_{\ell}$ and $t_{\ell}t_{h}$ and  for dilepton, they are $(Z/\gamma^{*})_\ell$ and $t_{\ell}t_{\ell}$. These distributions are obtained after applying the following cuts at the parton level: for monolepton $p_{T} (\ell) > 200$ GeV and for dilepton $M(\ell_1,\ell_2) > 200$ GeV.}
    \label{dist}
\end{figure*}

\begin{table*}
\caption{Cutflow for various background and signal contributions in the monolepton and dilepton final states. We present the number of events surviving the cuts in Table~\ref{tab:Cuts} for the HL-LHC ($14$~TeV, $3$~ab$^{-1}$; in parentheses, we show the corresponding numbers for vLQs). Two benchmark points are considered to highlight the dominance of LQ IP at higher LQ masses. \label{tab:nof}}
\centering
\renewcommand\baselinestretch{1.4}\selectfont
\begin{tabular*}{\textwidth}{@{\extracolsep{\fill}} lcccc}
\hline
\multirow{2}{*}{\textbf{Monolepton final state}} & \multicolumn{4}{c}{Selection Cuts} \\ \cline{2-5} 
                                        & C1         & C2                   & C3       & C4       \\ \hline \hline
Signal benchmarks                       &            &           &       &          \\ \hline
$M_{S_1 (U_1)}=1500$ GeV, $M_{\tau_{2}}= 500$ GeV (LQ PP)                          &   \multicolumn{1}{r}{$73$ ($526$)}         &      \multicolumn{1}{r}{$42$ ($304$)}                &   \multicolumn{1}{r}{$24$ ($178$)}       &   \multicolumn{1}{r}{$16$ ($114$)}       \\
$M_{S_1 (U_1)}=1500$ GeV, $M_{\tau_{2}}= 500$ GeV (LQ SP)                          &   \multicolumn{1}{r}{$1759$ ($3042$)}         &       \multicolumn{1}{r}{$1023$ ($1769$)}               &    \multicolumn{1}{r}{$621$ ($1073$)}      &   \multicolumn{1}{r}{$344$ ($595$)}       \\
$M_{S_1 (U_1)}=1500$ GeV, $M_{\tau_{2}}= 500$ GeV (LQ IP)                          &   \multicolumn{1}{r}{$1655$ ($6683$)}         &        \multicolumn{1}{r}{$1067$ ($4309$)}              &    \multicolumn{1}{r}{$765$ ($3091$)}      &    \multicolumn{1}{r}{$296$ ($1194$)}      \\ \hline
                                        &             \multicolumn{3}{l}{Total number of signal events:}       &   \multicolumn{1}{r}{$656$ ($1903$)}       \\ \hline
$M_{S_1 (U_1)}=2500$ GeV, $M_{\tau_{2}}= 500$ GeV (LQ PP)                          &   \multicolumn{1}{r}{$<1$ ($2$)}         &       \multicolumn{1}{r}{$<1$ ($1$)}               &    \multicolumn{1}{r}{$<1$ ($<1$)}      &   \multicolumn{1}{r}{$<1$ ($<1$)}       \\
$M_{S_1 (U_1)}=2500$ GeV, $M_{\tau_{2}}= 500$ GeV (LQ SP)                          &   \multicolumn{1}{r}{$79$ ($76$)}         &        \multicolumn{1}{r}{$45$ ($43$)}              &    \multicolumn{1}{r}{$24$ ($23$)}      &    \multicolumn{1}{r}{$17$ ($16$)}      \\
$M_{S_1 (U_1)}=2500$ GeV, $M_{\tau_{2}}= 500$ GeV (LQ IP)                          &    \multicolumn{1}{r}{$912$ ($1717$)}        &      \multicolumn{1}{r}{$575$ ($1084$)}                &     \multicolumn{1}{r}{$416$ ($785$)}     &   \multicolumn{1}{r}{$158$ ($297$)}       \\ \hline
                                        &            \multicolumn{3}{l}{ Total number of signal events:}        &   \multicolumn{1}{r}{$175$ ($313$)}       \\ \hline
Background processes                      &            &                      &          &          \\ \hline
$W_{\ell} (+2 j)$                                       & \multicolumn{1}{r}{$2.48 \times 10^7$}  &  \multicolumn{1}{r}{$8.38 \times 10^6$}  & \multicolumn{1}{r}{$1.65 \times 10^6$}  & \multicolumn{1}{r}{$131399$}  \\
$t_{\ell}t_{h} (+2 j)$                                      & \multicolumn{1}{r}{$1.01\times 10^6$}     &   \multicolumn{1}{r}{$658690$}            &   \multicolumn{1}{r}{$132560$}   &    \multicolumn{1}{r}{$2075$}   \\
$W_{\ell}W_{h} (+2 j)$                                      & \multicolumn{1}{r}{$457668$}      &    \multicolumn{1}{r}{$292473$}            &   \multicolumn{1}{r}{$45188$}   &    \multicolumn{1}{r}{$1350$}      \\
$W_{\ell}Z_{h} (+2 j)$                                      & \multicolumn{1}{r}{$160601$}      &    \multicolumn{1}{r}{$93678$}            &   \multicolumn{1}{r}{$13613$}   &    \multicolumn{1}{r}{$530$}      \\
$t_{\ell}b$                                      &   \multicolumn{1}{r}{$57961$}    &        \multicolumn{1}{r}{$15743$}              &      \multicolumn{1}{r}{$3786$}    &  \multicolumn{1}{r}{$57$}        \\
$t_{\ell}W_{h}+t_{h}W_{\ell}$                                      &  \multicolumn{1}{r}{$94245$}     &  \multicolumn{1}{r}{$65568$}              &  \multicolumn{1}{r}{$13690$}   &      \multicolumn{1}{r}{$12$}    \\
$t_{\ell}j$                                      &    \multicolumn{1}{r}{$2888$}        &         \multicolumn{1}{r}{$716$}             &      \multicolumn{1}{r}{$169$}    &     \multicolumn{1}{r}{$6$}     \\ \hline
                                        &           \multicolumn{3}{l}{ Total number of background events:}     &   \multicolumn{1}{r}{$135429$}       \\ \hline
\multirow{2}{*}{\textbf{Dilepton final state}}   & \multicolumn{4}{c}{Selection Cuts} \\ \cline{2-5} 
                                        & C1         & C2                   & C3       & C4       \\ \hline \hline
Signal benchmarks                       &            &                      &          &          \\ \hline
$M_{S_1 (U_1)}=1500$ GeV, $M_{\tau_{2}}= 500$ GeV (LQ PP)                          &    \multicolumn{1}{r}{$9$ ($43$)}        &    \multicolumn{1}{r}{$9$ ($43$)}                  &       \multicolumn{1}{r}{$4$ ($19$)}   &   \multicolumn{1}{r}{$2$ ($9$)}       \\
$M_{S_1 (U_1)}=1500$ GeV, $M_{\tau_{2}}= 500$ GeV (LQ SP)                          &    \multicolumn{1}{r}{$209$ ($362$)}        &           \multicolumn{1}{r}{$204$ ($354$)}           &   \multicolumn{1}{r}{$84$ ($146$)}       &   \multicolumn{1}{r}{$31$ ($53$)}       \\
$M_{S_1 (U_1)}=1500$ GeV, $M_{\tau_{2}}= 500$ GeV (LQ IP)                          &    \multicolumn{1}{r}{$72$ ($535$)}        &           \multicolumn{1}{r}{$40$ ($288$)}           &   \multicolumn{1}{r}{$17$ ($110$)}       &    \multicolumn{1}{r}{$3$ ($16$)}      \\ \hline
                                        &           \multicolumn{3}{l}{ Total number of signal events: }        &   \multicolumn{1}{r}{$36$ ($78$)}       \\ \hline
$M_{S_1 (U_1)}=2500$ GeV, $M_{\tau_{2}}= 500$ GeV (LQ PP)                          &      \multicolumn{1}{r}{$<1$ ($<1$)}      &              \multicolumn{1}{r}{$<1$ 
($<1$)}      &   \multicolumn{1}{r}{$<1$ ($<1$)}       &   \multicolumn{1}{r}{$<1$ ($<1$)}       \\
$M_{S_1 (U_1)}=2500$ GeV, $M_{\tau_{2}}= 500$ GeV (LQ SP)                          &     \multicolumn{1}{r}{$10$ ($10$)}       &      \multicolumn{1}{r}{$9$ ($9$)}                &    \multicolumn{1}{r}{$5$ ($5$)}      &    \multicolumn{1}{r}{$3$ ($3$)}      \\
$M_{S_1 (U_1)}=2500$ GeV, $M_{\tau_{2}}= 500$ GeV (LQ IP)                          &      \multicolumn{1}{r}{$133$ ($136$)}      &      \multicolumn{1}{r}{$71$ ($76$)}                &    \multicolumn{1}{r}{$27$ ($37$)}      &   \multicolumn{1}{r}{$4$ ($7$)}       \\ \hline
                                        &           \multicolumn{3}{l}{ Total number of signal events:}        &  \multicolumn{1}{r}{$7$ ($10$)}        \\ \hline
Background processes                      &            &                      &          &          \\ \hline
$(Z/\gamma^{*})_{\ell} (+2 j)$                                       &  \multicolumn{1}{r}{$6.08 \times 10^{6}$}          &    \multicolumn{1}{r}{$1.19\times 10^{6}$}                  &   \multicolumn{1}{r}{$285621$}       &  \multicolumn{1}{r}{$368$}        \\
$t_{\ell}t_{\ell} (+2 j)$                                      &     \multicolumn{1}{r}{$1.22\times 10^{6}$}       &    \multicolumn{1}{r}{$318906$}                  &      \multicolumn{1}{r}{$35356$}    &   \multicolumn{1}{r}{$296$}       \\
$W_{\ell}W_{\ell} (+2 j)$                                      &      \multicolumn{1}{r}{$332129$}      &      \multicolumn{1}{r}{$95628$}                &      \multicolumn{1}{r}{$16737$}    &    \multicolumn{1}{r}{$290$}      \\
$t_{\ell}W_{\ell}$                                     &      \multicolumn{1}{r}{$161677$}      &    \multicolumn{1}{r}{$26615$}                  &     \multicolumn{1}{r}{$3965$}     &    \multicolumn{1}{r}{$16$}      \\
$W_{h}Z_{\ell} (+2 j)$                                      &   \multicolumn{1}{r}{$1670$}         &     \multicolumn{1}{r}{$592$}                 &      \multicolumn{1}{r}{$147$}    &    \multicolumn{1}{r}{$<1$}      \\
$Z_{\ell}Z_{h} (+2 j)$                                      &    \multicolumn{1}{r}{$200$}        &       \multicolumn{1}{r}{$41$}               &      \multicolumn{1}{r}{$5$}    &    \multicolumn{1}{r}{$<1$}      \\ \hline
                                        &            \multicolumn{3}{l}{ Total number of background events: }     &   \multicolumn{1}{r}{$970$}       \\ \hline
\end{tabular*}
\end{table*}

\subsubsection{Monolepton}
\noindent
A monolepton final state ($\ell = e,\mu$) can arise from the decay of $\tau_2$ pairs produced through different LQ production modes. In our analysis, we consider the following two decay chains. 
\begin{align*}
pp \to 
\left\{
\begin{array}{l}
\ell_q \ell_q \\
\ell_q\, \tau_2\ (+j) \\
\tau_2 \tau_2\ (+j)
\end{array}
\right\}
&\to 
\left\{
\begin{array}{l}
(j\,\tau_2)(j\,\tau_2) \\
(j\,\tau_2)\, \tau_2\ (+j) \\
\tau_2 \tau_2\ (+j)
\end{array}
\right\}\nn\\\to&\left\{\begin{array}{l}
    \nu_{\tau} W_{h}^{\mp}\, \nu_{\tau} W_{\ell}^{\pm} + \mbox{jet(s)}\\
    \tau_{\ell} Z_{h}\, \nu_{\ell} W_{h}^{\pm}+ \mbox{jet(s)}\end{array}\right\}. 
\end{align*}

In the first chain, both $\tau_2$ decay into $W + \nu$, with one $W$ decaying leptonically ($W_\ell$) and the other hadronically ($W_h$). In the second scenario, one $\tau_2$ decays into $W + \nu$, while the other decays into $Z + \tau$. A monolepton final state is obtained if, either the $\tau$ decays leptonically ($\tau_\ell$) and the $W$ hadronically, or the $\tau$ decays hadronically ($\tau_h$) and the $W$ leptonically, with the $Z$ decaying hadronically ($Z_h$) in both cases. Other decay chains can, in principle, contribute to monolepton final states, but their cross sections are negligible and hence they are not included in this study.

\subsubsection{Dilepton}
\noindent
In the dilepton channel, we combine the $ee$, $\mu\mu$, and $e\mu$ modes. These arise primarily from two dominant decay chains.
\begin{align*}
pp \to 
\left\{
\begin{array}{l}
\ell_q \ell_q \\
\ell_q\, \tau_2\ (+j) \\
\tau_2 \tau_2\ (+j)
\end{array}
\right\}
&\to 
\left\{
\begin{array}{l}
(j\,\tau_2)(j\,\tau_2) \\
(j\,\tau_2)\, \tau_2\ (+j) \\
\tau_2 \tau_2\ (+j)
\end{array}
\right\}\nn\\\to&\left\{\begin{array}{l}
    \nu_{\tau_2} W_{\ell}^{\mp}\, \nu_{\tau_2} W_{\ell}^{\pm} + \mbox{jet(s)}\\
    \nu_{\tau_2} W_{\ell}^{\mp}\, \tau_{\ell}^{\mp} Z_{h}+ \mbox{jet(s)}\end{array}\right\}.
\end{align*}
In the first case, both $\tau_2$ decay into $W + \nu$, with both $W$ bosons subsequently decaying leptonically, yielding two charged leptons and missing energy. In the second case, one $\tau_2$ decays into $W + \nu$ while the other decays into $Z + \tau$. A dilepton final state is obtained when both the $W$ and the $\tau$ decay leptonically, with the $Z$ decaying hadronically. Other possible decay chains of $\tau_2$ pairs can also lead to dilepton signatures, but their cross sections are negligible and are therefore not considered in this study.

\subsubsection{Trilepton}
\noindent
For trilepton final states, both $\tau_2$ can decay either symmetrically into $Z + \nu$ or asymmetrically, with one decaying into $W + \nu$ and the other into $Z + \nu$. The relevant production and decay chains are as follows.
\begin{align*}
pp \to 
\left\{
\begin{array}{l}
\ell_q \ell_q \\
\ell_q\, \tau_2\ (+j) \\
\tau_2 \tau_2\ (+j)
\end{array}
\right\}
&\to 
\left\{
\begin{array}{l}
(j\,\tau_2)(j\,\tau_2) \\
(j\,\tau_2)\, \tau_2\ (+j) \\
\tau_2 \tau_2\ (+j)
\end{array}
\right\} \\
&\to~ 
\nu_{\tau_2} W_{\ell}^{\mp}\, \tau_{h}^{\mp} Z_{\ell} + \text{jet(s)}.
\end{align*}

Although the cross sections for these channels are relatively small, resulting in fewer signal events, the corresponding SM background is also significantly lower. This suppression of background can enhance the sensitivity of the analysis in the trilepton channel, making it a valuable probe despite the limited signal yield. Several previous studies have explored trilepton signatures in related contexts (see Refs.~\cite{Kang16,Accomando:2017qcs,Helo:2018rll}).

\subsubsection{Quadlepton}
\noindent
The quadlepton final states, consisting of four charged leptons, can originate either from a symmetric decay, where both $\tau_2$ decay into $Z + \tau$, or from an asymmetric configuration in which one $\tau_2$ decays into $W + \nu$ and the other into $Z + \tau$. The relevant production and decay chains are as follows.
\begin{align*}
pp \to 
\left\{
\begin{array}{l}
\ell_q \ell_q \\
\ell_q\, \tau_2\ (+j) \\
\tau_2 \tau_2\ (+j)
\end{array}
\right\}
&\to 
\left\{
\begin{array}{l}
(j\,\tau_2)(j\,\tau_2) \\
(j\,\tau_2)\, \tau_2\ (+j) \\
\tau_2 \tau_2\ (+j)
\end{array}
\right\}\nn\\\to&\left\{\begin{array}{l}
    \tau_{h}^{\mp} Z_{\ell}\, \tau_{h}^{\mp} Z_{\ell} + \mbox{jet(s)}\\
    \nu_{\tau_2} W_{\ell}^{\mp}\, \tau_{\ell}^{\mp} Z_{\ell}+ \mbox{jet(s)}\end{array}\right\}.
\end{align*}

Although the event rate for quadlepton final states is low—primarily due to the requirement of multiple leptonic decays of gauge bosons—the corresponding SM background is also highly suppressed. Consequently, despite being statistically limited, this channel offers a particularly clean environment and can provide strong sensitivity to new physics signatures involving LQ-induced $\tau_2$ decays. Similar quadlepton signatures have also been studied in the context of other models (see Ref.~\cite{Huitu08}).

\subsubsection{Five and six leptons}
\noindent
Final states with charged-lepton multiplicity greater than four are strongly suppressed by their small cross sections and are therefore extremely challenging to observe at colliders. Five-lepton final states can arise exclusively from the symmetric decay of both $\tau_2$ particles via the $Z + \tau$ channel. The representative production and decay chain is given below.
\begin{align*}
pp \to 
\left\{
\begin{array}{l}
\ell_q \ell_q \\
\ell_q\, \tau_2\ (+j) \\
\tau_2 \tau_2\ (+j)
\end{array}
\right\}
&\to 
\left\{
\begin{array}{l}
(j\,\tau_2)(j\,\tau_2) \\
(j\,\tau_2)\, \tau_2\ (+j) \\
\tau_2 \tau_2\ (+j)
\end{array}
\right\} \\
&\to~ 
\tau_{h}^{\mp} Z_{\ell}\, \tau_{\ell}^{\mp} Z_{\ell} + \text{jet(s)}.
\end{align*}

Similar to the quadlepton case, the five-lepton channel suffers from a very low event yield due to the requirement of multiple leptonic decays. However, the corresponding SM background for such high-multiplicity final states is also extremely suppressed. Despite the limited statistics, this channel can still serve as a clean and promising probe of BSM scenarios.

Six-lepton final states can originate from the fully symmetric decay of both $\tau_2$ particles via the $Z + \tau$ channel, with all subsequent decays occurring leptonically. The full production and decay sequence is described below.
\begin{align*}
pp \to 
\left\{
\begin{array}{l}
\ell_q \ell_q \\
\ell_q\, \tau_2\ (+j) \\
\tau_2 \tau_2\ (+j)
\end{array}
\right\}
&\to 
\left\{
\begin{array}{l}
(j\,\tau_2)(j\,\tau_2) \\
(j\,\tau_2)\, \tau_2\ (+j) \\
\tau_2 \tau_2\ (+j)
\end{array}
\right\} \\
&\to~ 
\tau_{\ell}^{\mp} Z_{\ell}\, \tau_{\ell}^{\mp} Z_{\ell} + \text{jet(s)}.
\end{align*}
This channel is expected to be highly suppressed due to the small branching ratios associated with multiple leptonic decays. Nevertheless, the SM background for six-lepton final states is practically negligible. As a result, this channel offers an exceptionally clean signature and may serve as a compelling probe of new physics.

\subsubsection{Fatjet plus missing energy}
\noindent
In this scenario, the $\tau_2$ particles may decay either symmetrically or asymmetrically, with all subsequent decay products undergoing hadronic decays. This results in a final state characterised by one or more fat jets accompanied by missing transverse energy, but no isolated charged leptons. The relevant production and decay processes are outlined below.
\begin{align*}
pp \to 
\left\{
\begin{array}{l}
\ell_q \ell_q \\
\ell_q\, \tau_2\ (+j) \\
\tau_2 \tau_2\ (+j)
\end{array}
\right\}
&\to 
\left\{
\begin{array}{l}
(j\,\tau_2)(j\,\tau_2) \\
(j\,\tau_2)\, \tau_2\ (+j) \\
\tau_2 \tau_2\ (+j)
\end{array}
\right\}\nn\\\to&\left\{\begin{array}{l}
    \tau_{h}^{\mp} Z_{h}\, \tau_{h}^{\mp} Z_{h} + \mbox{jet(s)}\\
    \nu_{\tau} W_{h}^{\mp}\, \tau_{h}^{\mp} Z_{h}+ \mbox{jet(s)}\end{array}\right\}
\end{align*}

This channel benefits from a relatively higher signal rate due to the large hadronic branching fractions. However, the SM background—dominated by QCD multijet and $W/Z$+jets processes—is also expected to be substantial. The distinctive kinematic features of fat jets and missing energy provide effective handles for background suppression, making this an important complementary search channel for new physics involving LQ-mediated $\tau_2$ production. Similar final states have been studied in the context of the Inert Higgs Doublet Model (see Ref.~\cite{Bhardwaj:2019mts}).

\subsubsection{Displaced vertex}
\noindent
The VLL can also give rise to displaced vertex signatures if its decay width is sufficiently small. The decay width is controlled by the off-diagonal term in the mass matrix. In this work, we assume this term to be of the order of the SM $\tau$ lepton mass. However, if it were instead of the order of the electron mass, the $\tau_2$ could produce displaced vertices. Studies of long-lived VLLs in this context can be found in Refs.~\cite{Bernreuther:2023uxh,Bandyopadhyay:2023joz,Cao:2023smj}. For even smaller decay widths, the VLL would decay outside the detector, leading to missing-energy signatures. Such scenarios demand dedicated search strategies at the LHC.

\section{Signal, Background and Selection criteria} \label{signal}
\noindent
The signal regions are categorised into monolepton and dilepton final states. The monolepton channel consists of events containing either an electron or a muon, while the dilepton channel includes both same-flavour (electron–electron or muon–muon) and mixed-flavour (electron–muon) pairs. In both channels, the heavy vector-like lepton $\tau_2$ may decay either symmetrically into $W\nu$ or asymmetrically into $W\nu$ and $Z\tau$.

The dominant SM backgrounds in the monolepton channel arise from $W_\ell$+jets and $t_\ell t_h$+jets, whereas in the dilepton channel they originate primarily from $Z_\ell$+jets and $t_\ell t_\ell$+jets processes. All relevant backgrounds are generated with parton-level cuts applied: for the monolepton channel, we impose $\slashed{E}_T > 250$ GeV, while for the dilepton channel, we require $M(\ell_1,\ell_2) > 200$ GeV.
 To suppress these backgrounds, events are required to contain at least one AK4 jet and at least one fatjet in both signal regions. In addition, a $b$-veto is imposed in both the mono- and dilepton channels to further reduce the large $t\bar{t}$ background. The complete set of selection cuts, guided by discriminating kinematic distributions between signal and background, is presented in Table~\ref{tab:Cuts}, with the corresponding distributions shown in Fig.~\ref{dist}.

For the monolepton channel, the selection strategy is optimised for the IP topology, which dominates at higher LQ masses and larger coupling strengths. In the lower LQ mass region, where PP contributions dominate, one could, in principle, optimise the selection cuts using PP kinematics. However, we find that the IP-optimised cuts also perform well in this regime, providing a model-independent reach comparable to that achievable with PP-optimised cuts. In contrast, for the dilepton channel, IP-optimised cuts—while offering good sensitivity in the high-mass region—fail to provide adequate sensitivity in the low-mass regime. Consequently, they cannot be employed effectively across the entire mass range as the IP-optimised cuts significantly weaken the sensitivity in the low-mass region dominated by PP, mainly due to certain jet-related requirements specific to the IP topology. To ensure robust, model-independent sensitivity, we therefore adopt PP-optimised selection cuts across the entire mass range in the dilepton channel. These cuts not only yield stronger exclusion bounds at low masses but also maintain comparable sensitivity in the high-mass region that is achievable with IP-optimised cuts.

\begin{figure*}[]
    \centering
    \captionsetup[subfigure]{labelformat=empty}\vspace{-0.4cm}
    \subfloat[(a)]{\includegraphics[width=0.2\textwidth]{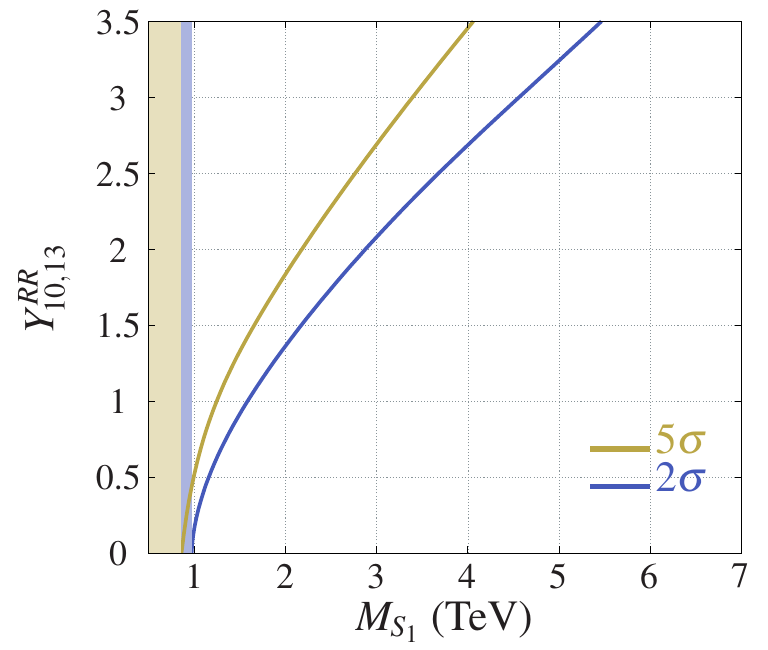}\label{S1_mono}}\hspace{0.5cm}
    \subfloat[(b)]{\includegraphics[width=0.2\textwidth]{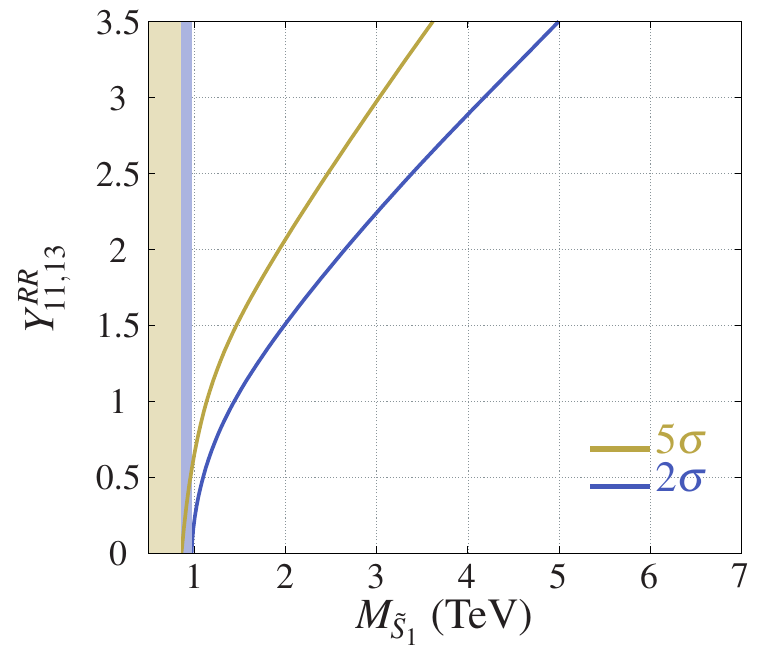}\label{S1bar_mono}}\hspace{0.5cm}
    \subfloat[(c)]{\includegraphics[width=0.2\textwidth]{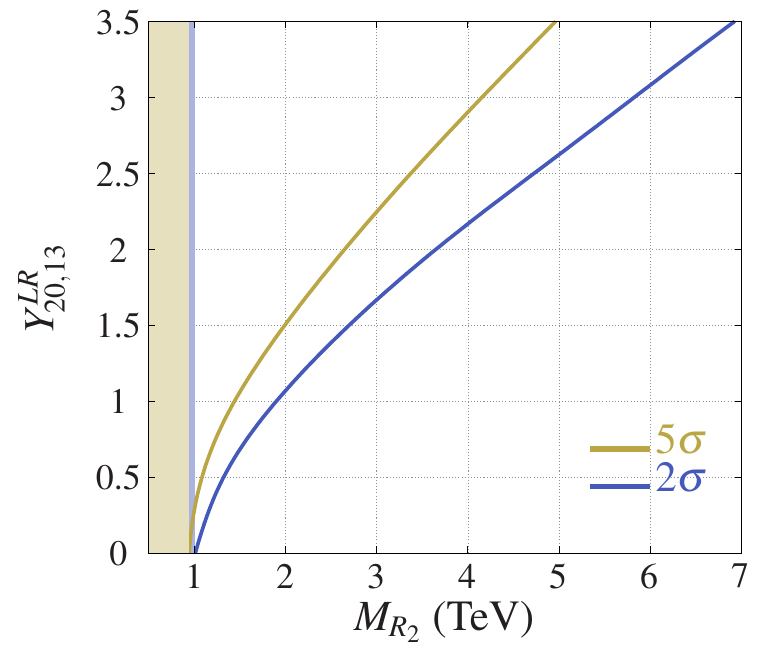}\label{R2t_mono}}\hspace{0.5cm}
    \subfloat[(d)]{\includegraphics[width=0.2\textwidth]{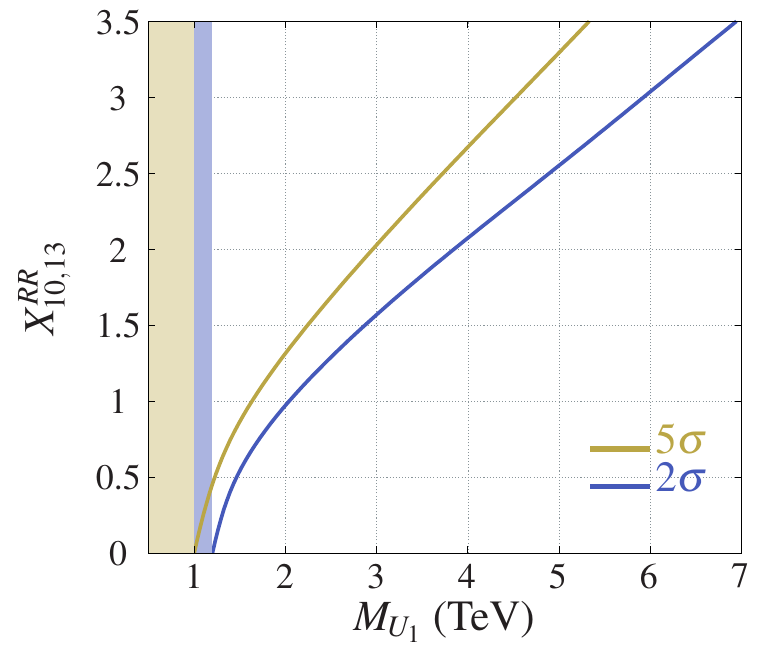}\label{U1k0_mono}}\\\vspace{-0.4cm}
    \subfloat[(e)]{\includegraphics[width=0.2\textwidth]{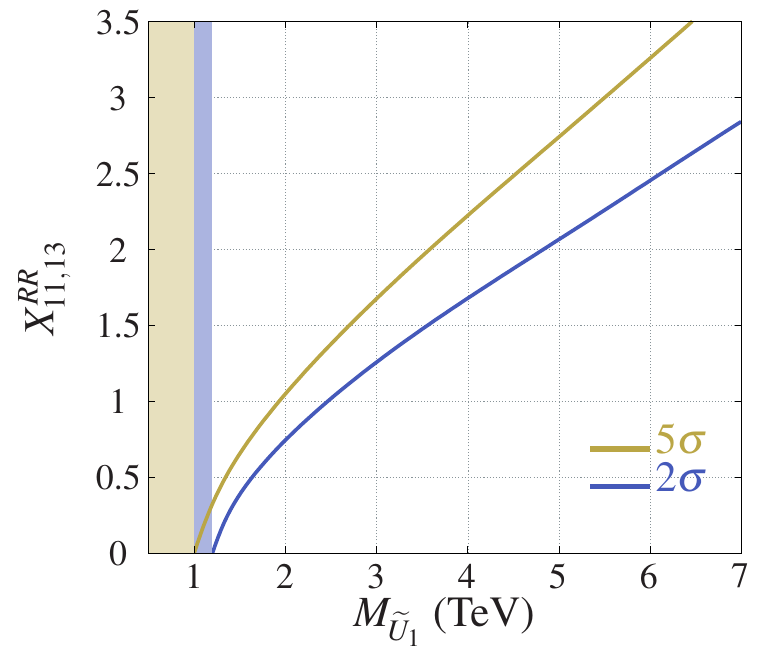}\label{U1bar_mono}}\hspace{0.5cm}
    \subfloat[(f)]{\includegraphics[width=0.2\textwidth]{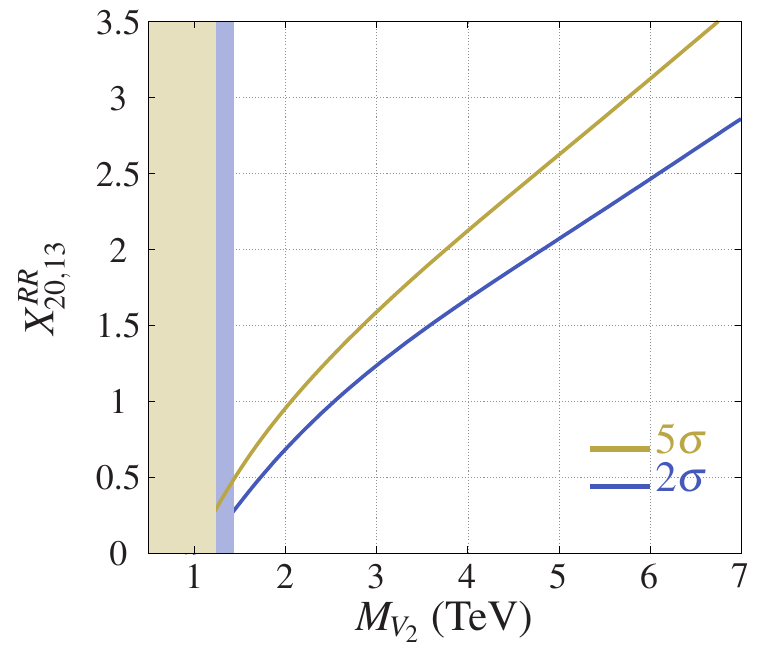}\label{V2t_mono}}\hspace{0.5cm}
    \subfloat[(g)]{\includegraphics[width=0.2\textwidth]{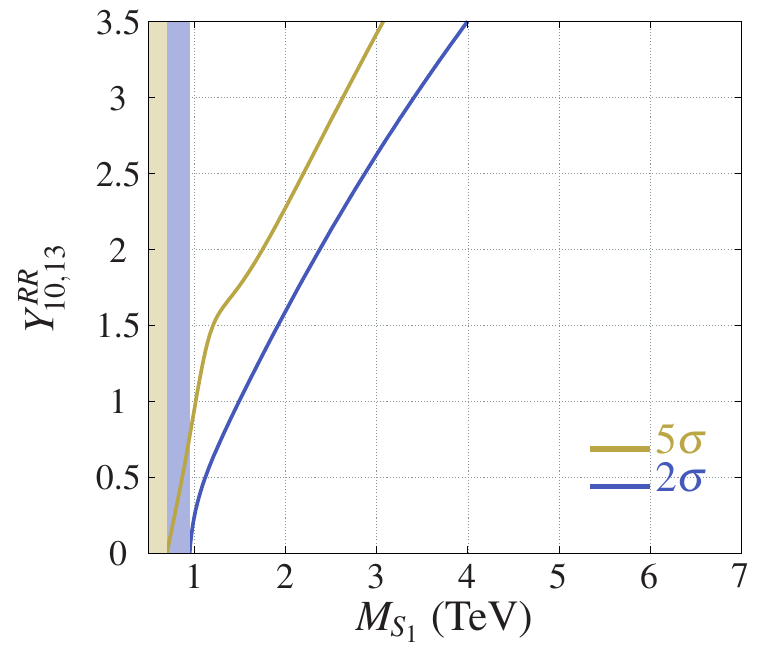}\label{S1_di}}\hspace{0.5cm}
    \subfloat[(h)]{\includegraphics[width=0.2\textwidth]{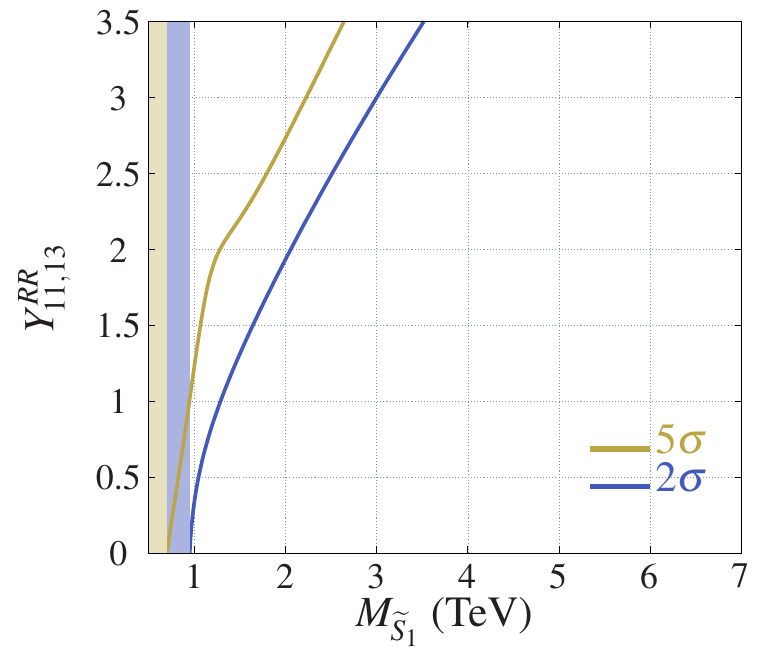}\label{S1bar_di}}\\\vspace{-0.4cm}
    \subfloat[(i)]{\includegraphics[width=0.2\textwidth]{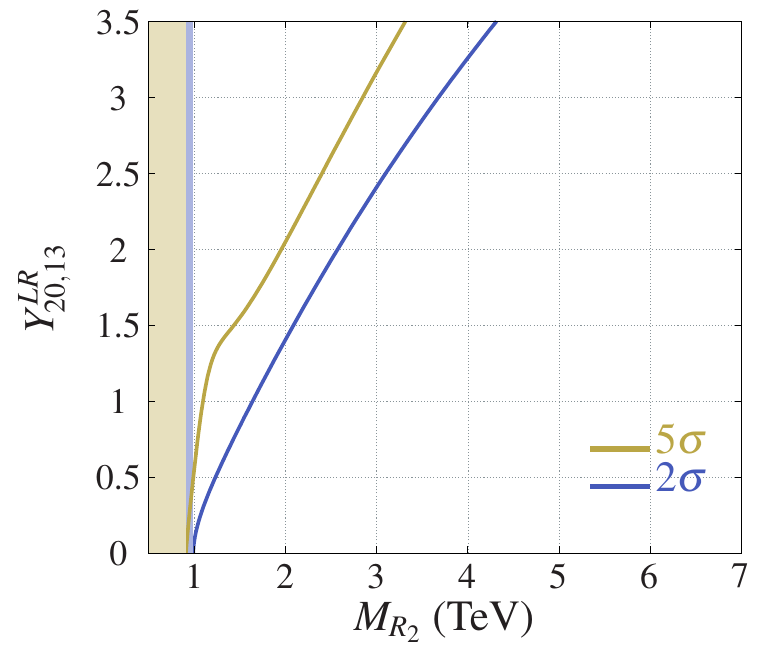}\label{R2t_di}}\hspace{0.5cm}
    \subfloat[(j)]{\includegraphics[width=0.2\textwidth]{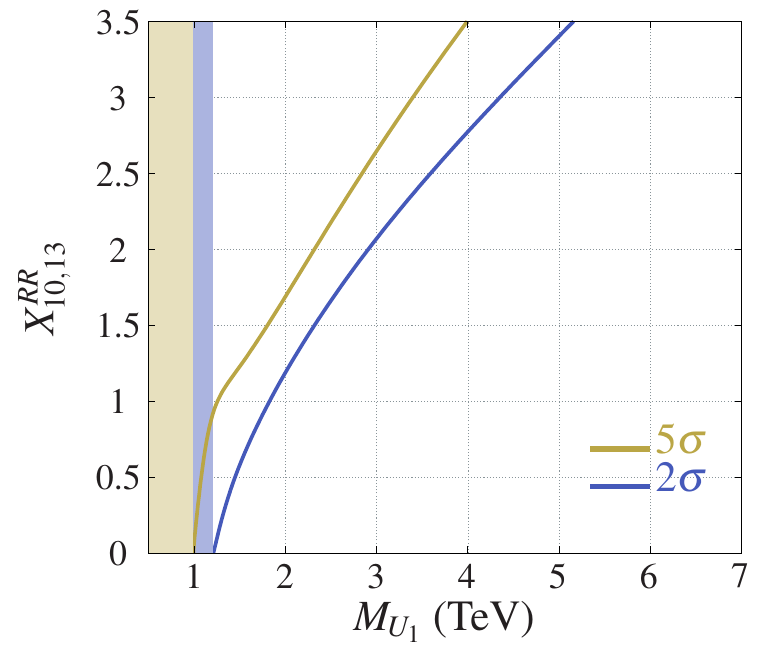}\label{U1_di}}\hspace{0.5cm}
    \subfloat[(k)]{\includegraphics[width=0.2\textwidth]{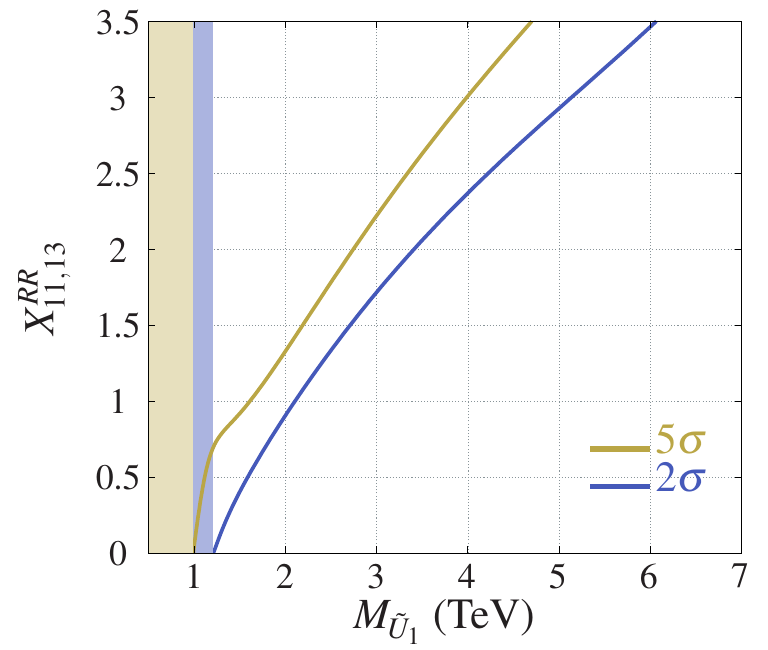}\label{U1bar_di}}\hspace{0.5cm}
    \subfloat[(l)]{\includegraphics[width=0.2\textwidth]{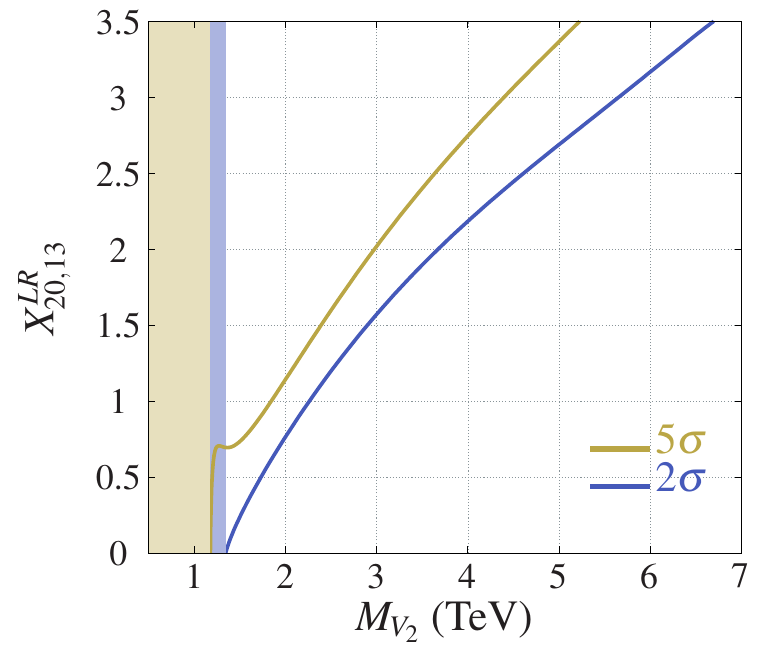}\label{V2t_di}}
    \caption{The $2\sigma$  and $5\sigma$ regions at the HL-LHC: [(a) -- (f)] for monolepton final states and [(g) -- (l)] for dilepton final states. The vertical shaded regions indicate the model-independent limits (i.e., obtained only with PP by setting $x/y\to 0$) for different LQs. All these limits are obtained for $M_{\tau_2} = 500$ GeV. We set $\kappa=1$ in the vLQ plots.}
    \label{MLQvslam}
    \subfloat[(a)]{\includegraphics[width=0.2\textwidth]{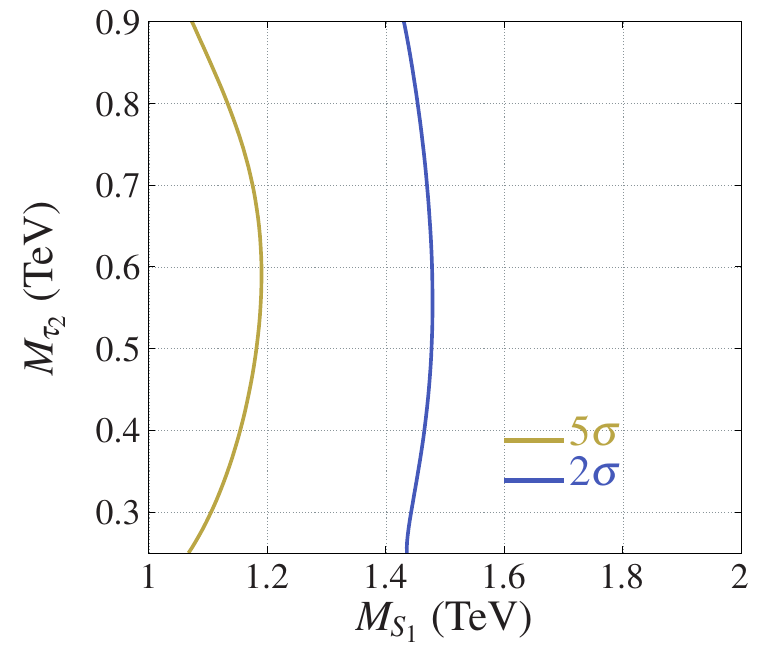}\label{S1_vll_mono}}\hspace{0.5cm}
    \subfloat[(b)]{\includegraphics[width=0.2\textwidth]{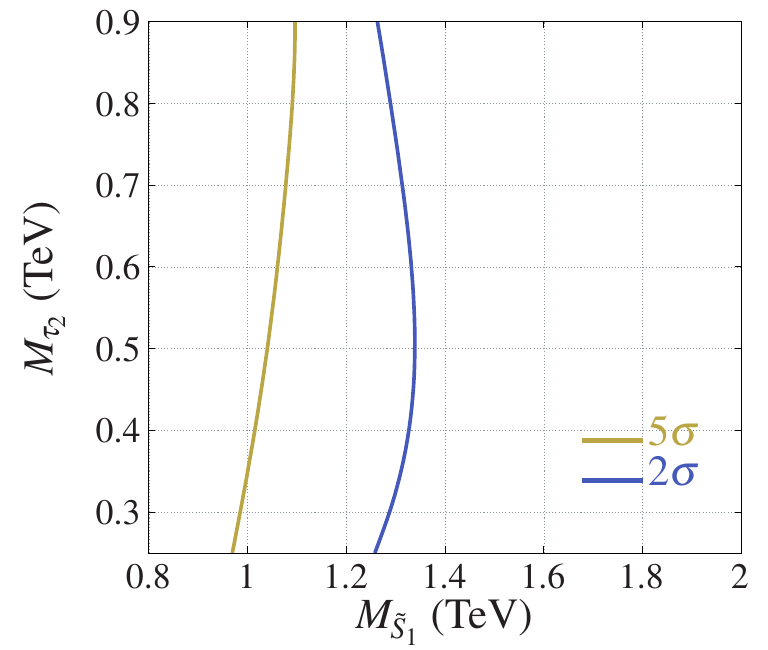}\label{S1t_vll_mono}}\hspace{0.5cm}
    \subfloat[(c)]{\includegraphics[width=0.2\textwidth]{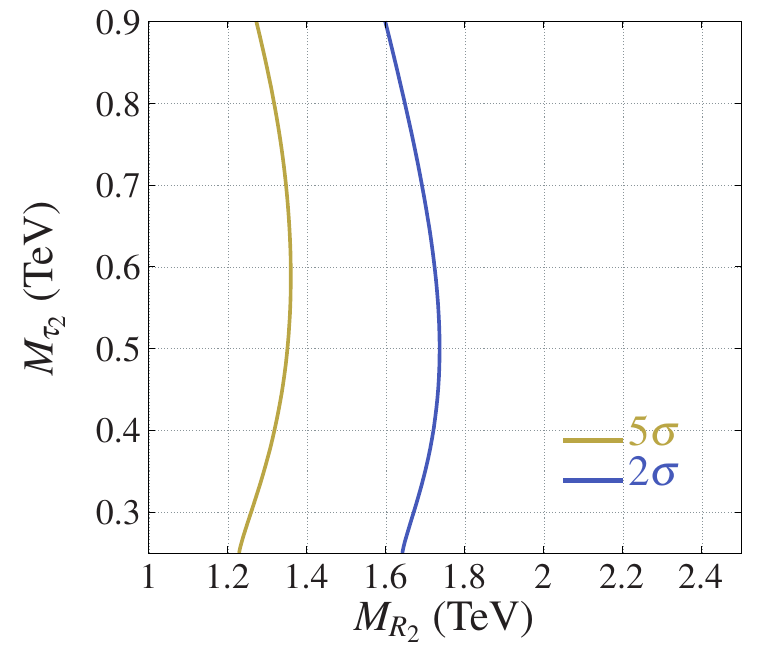}\label{R2t_vll_mono}}\hspace{0.5cm}
    \subfloat[(d)]{\includegraphics[width=0.2\textwidth]{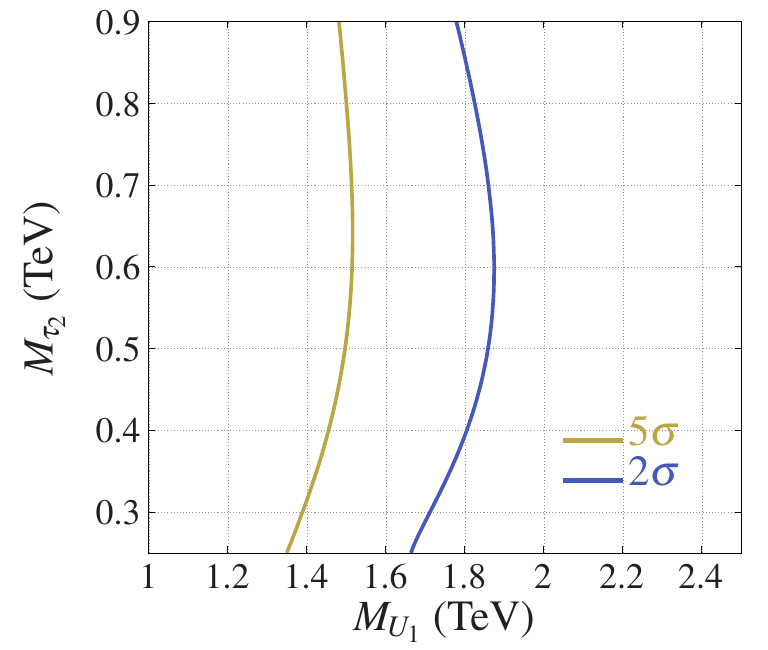}\label{S1_vll_mono}}\\\vspace{-0.4cm}
    \subfloat[(e)]{\includegraphics[width=0.2\textwidth]{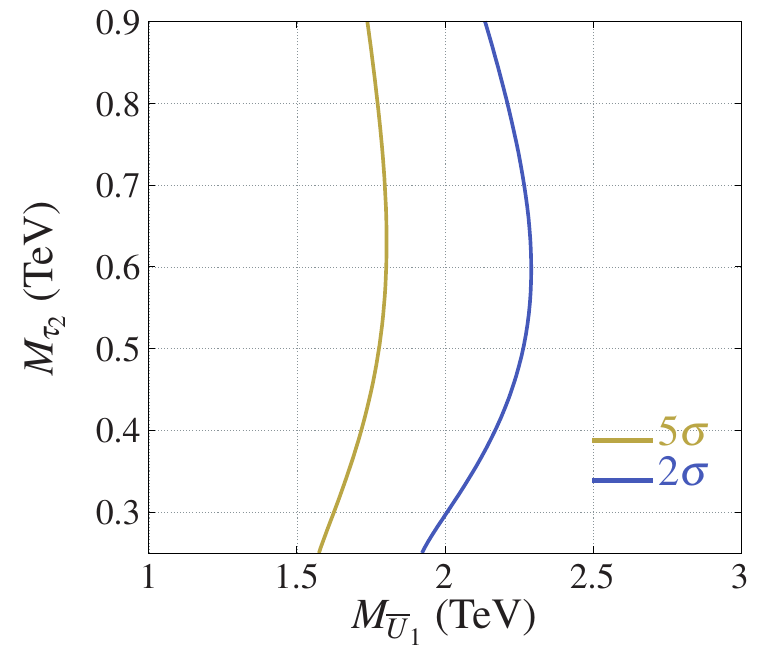}\label{S1bar_vll_mono}}\hspace{0.5cm}
    \subfloat[(f)]{\includegraphics[width=0.2\textwidth]{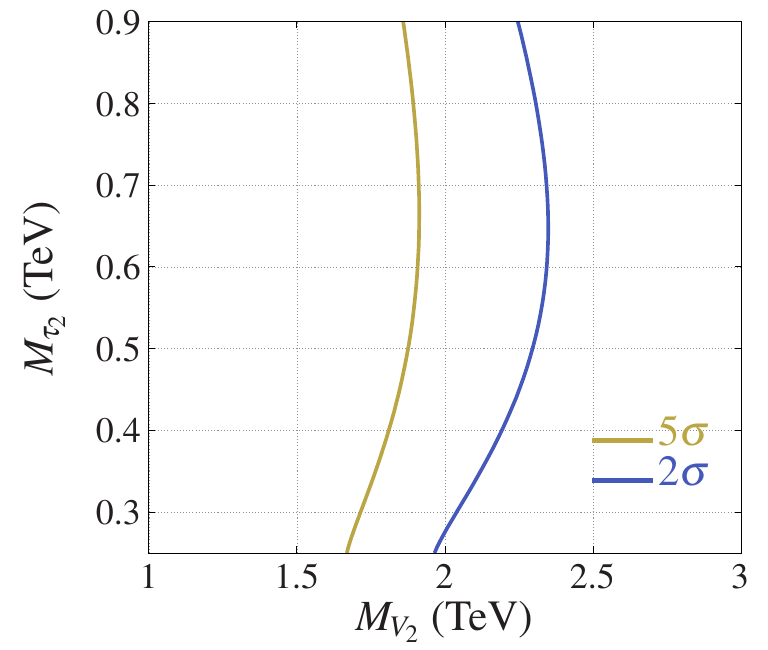}\label{V2_vll_mono}}\hspace{0.5cm}
    \subfloat[(g)]{\includegraphics[width=0.2\textwidth]{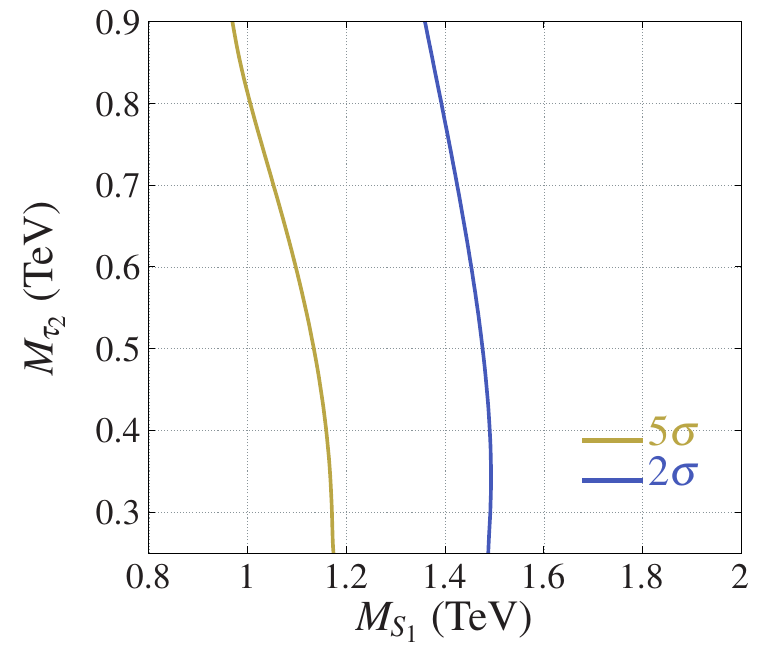}\label{S1_vll_di}}\hspace{0.5cm}
    \subfloat[(h)]{\includegraphics[width=0.2\textwidth]{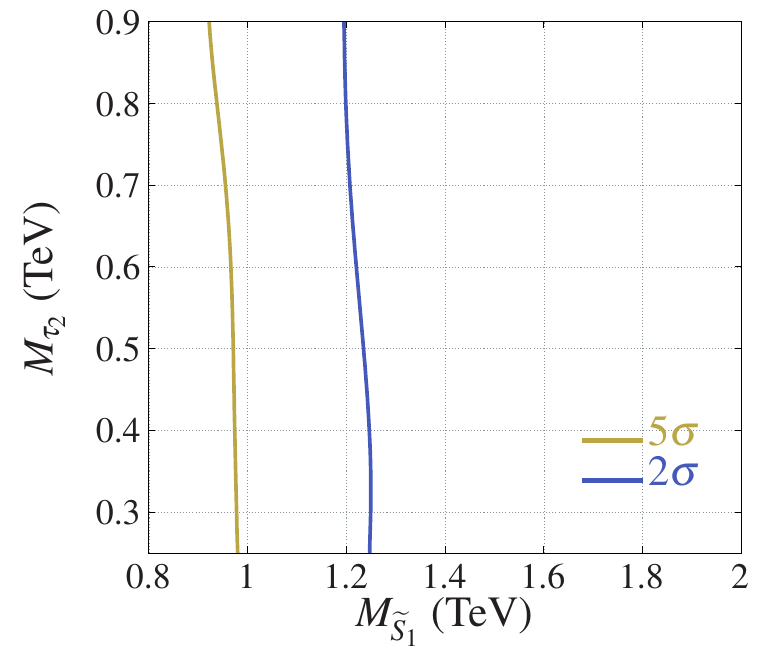}\label{S1t_vll_di}}\\\vspace{-0.4cm}
    \subfloat[(i)]{\includegraphics[width=0.2\textwidth]{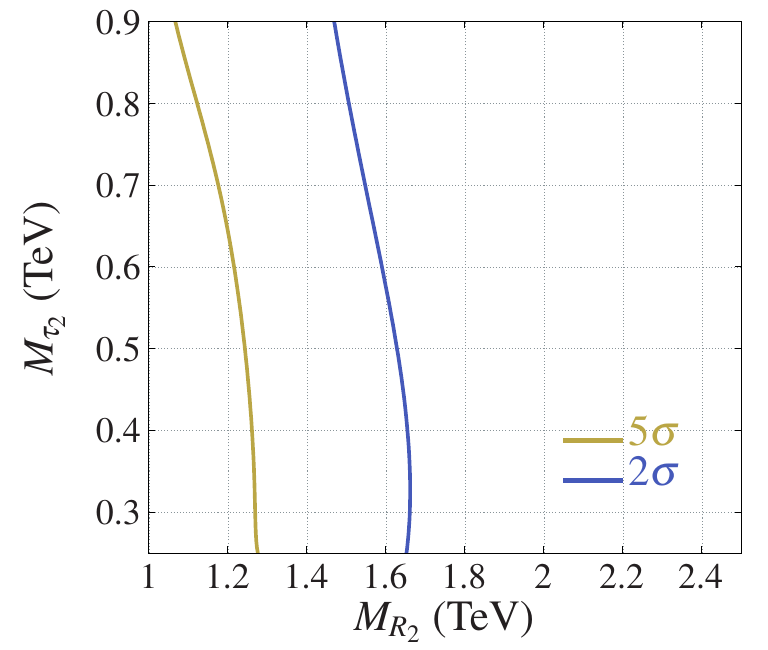}\label{R2_vll_di}}\hspace{0.5cm}
    \subfloat[(j)]{\includegraphics[width=0.2\textwidth]{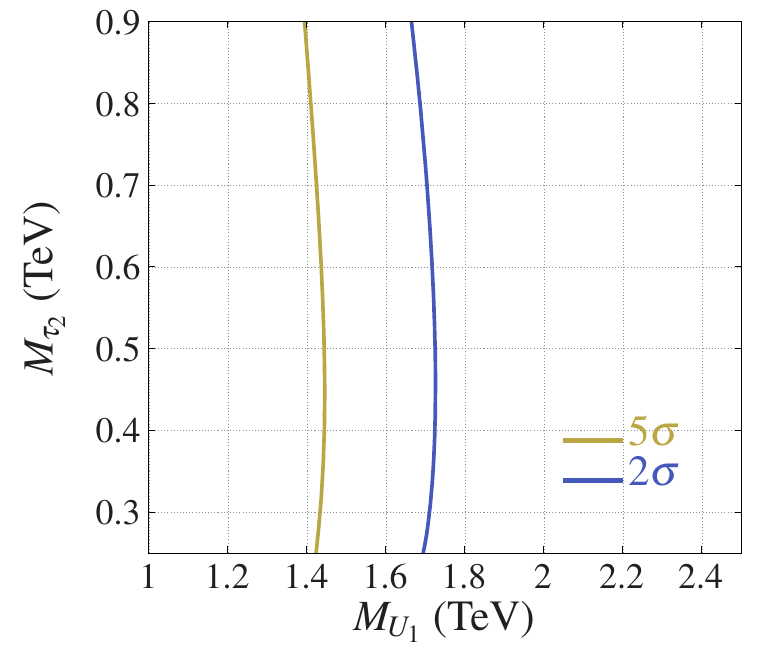}\label{U1_vll_di}}\hspace{0.5cm}
    \subfloat[(k)]{\includegraphics[width=0.2\textwidth]{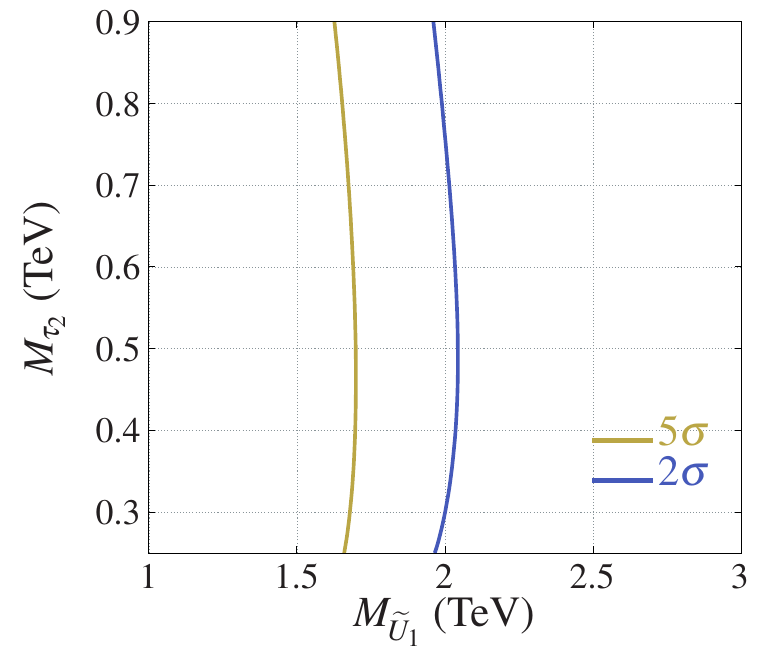}\label{U1t_vll_di}}\hspace{0.5cm}
    \subfloat[(l)]{\includegraphics[width=0.2\textwidth]{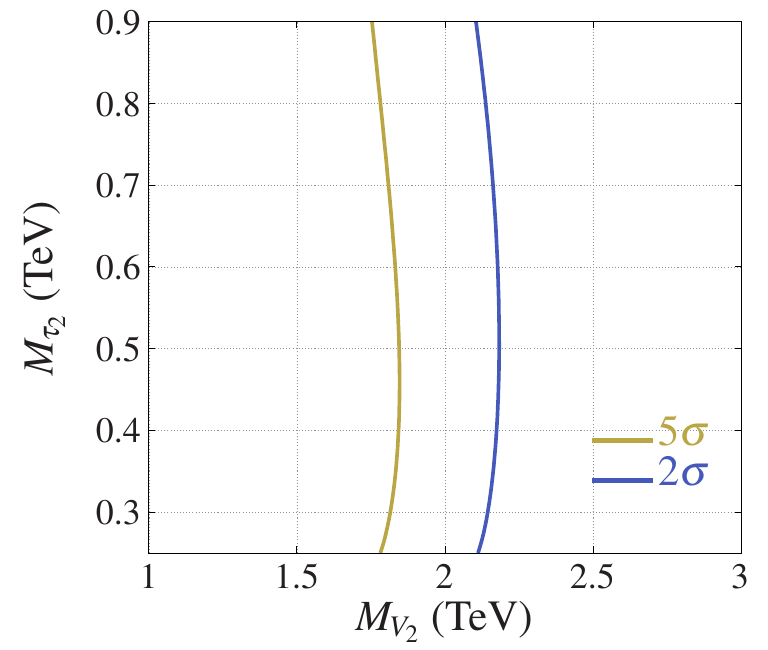}\label{V2_vll_di}}
    \caption{The $2\sigma$ and $5\sigma$ regions for the monolepton [(a)--(f)] and dilepton [(g)--(l)] final states on the $M_{\ell_q}-M_{\tau_2}$ planes with fixed coupling, $x/y = 1$.}\label{LQvsvll}
\end{figure*}

\section{HL-LHC prospects} 
\label{hllhc}
\noindent
We present the projected sensitivity of our model at the HL-LHC, operating at a centre-of-mass energy of $14$~TeV with an integrated luminosity of $\mathcal{L} = 3~\text{ab}^{-1}$. The number of signal and background events that survive the full set of selection cuts is presented in Table~\ref{tab:nof} for both monolepton and dilepton channels. The total number of signal events, coming from PP, SP, IP and II contributions, is computed using the expression:
\begin{equation}
N_{\text{S}} = \Big( \sigma_{\text{PP}} \cdot \epsilon_{\text{PP}} + \lambda^2 \cdot \sigma_{\text{SP}} \cdot \epsilon_{\text{SP}} +  \lambda^4 \cdot \sigma_{\text{IP}} \cdot \epsilon_{\text{IP}} +  \lambda^2 \cdot \sigma_{\text{II}} \cdot \epsilon_{\text{II}} \Big)\mathcal{L},
\end{equation}
where $\sigma_{xx}$ and $\epsilon_{xx}$ ($xx=$ PP, SP, IP, II) denote the production cross sections and the selection efficiencies of the respective signal topologies. 

To quantify the statistical significance of the signal, we use the Asimov approximation for the $\mathcal{Z}$-score (see Ref.~\cite{Cowan:2010js}):
\begin{equation}
\label{eq:Zscore}
\mathcal{Z} = \sqrt{2\left(N_{\text{S}} + N_{\text{B}}\right)\ln\left(\frac{N_{\text{S}} + N_{\text{B}}}{N_{\text{B}}}\right) - 2N_{\text{S}}}\,,
\end{equation}
where $N_{\text{S}}$ and $N_{\text{B}}$ represent the number of signal and background events, respectively.

The resulting $5\sigma$ discovery and $2\sigma$ exclusion contours in the $M_{\ell_q}$--$\lambda$ plane (for fixed $M_{\tau_2} = 500$~GeV) are shown in Fig.~\ref{MLQvslam} (see Ref.~\cite{Bhaskar:2024wic} for the nomenclature). The corresponding contour plots in the $M_{\text{LQ}}$--$M_{\tau_2}$ plane are shown in Fig.~\ref{LQvsvll}. We observe that exclusion and discovery sensitivities improve with increasing $M_{\tau_2}$ and reach a maximum around 500 GeV, indicating that the selection cuts are optimally tuned for this mass region. For higher $M_{\tau_2}$ values, the contours bend toward the lower $M_{\ell_q}$, as expected to maintain the number of signal events.

For the scalar LQs $S_1$ and $\widetilde{S}_1$, we find that the coupling-independent limits in both the monolepton and dilepton channels are nearly identical. At higher LQ masses, where IP dominates, the sensitivity to $S_1$ is slightly enhanced due to its coupling to up-type quarks, which results in a larger IP cross section. In contrast, $\widetilde{S}_1$ couples to down-type quarks, leading to a smaller cross section. Nevertheless, the overall difference remains small because the II contributions are destructive for $S_1$ but constructive for $\widetilde{S}_1$.
For vLQs $U_1$ and $\widetilde{U}_1$, the effect is more pronounced. The $U_1$ LQ couples to down-type quarks, and the nature of II is destructive. Whereas $\widetilde{U}_1$ couples to up-type quarks and the nature of II is constructive. As a result, the exclusion reach for $\widetilde{U}_1$ is significantly stronger compared to $U_1$.

Throughout this work, we have assumed the LQ couplings to SM quark-lepton pairs to be negligible. Here, we briefly comment on how the presence of additional decay modes, specifically LQ decays to a SM quark-lepton pair, could impact our results. If such decay modes occur with significant branching fractions, they can substantially contaminate the mono- and dilepton signal channels considered in this study. To keep this contamination under control, the corresponding decay couplings must be sufficiently small. For example, in the dilepton channel, for the case of constructively interfering $\widetilde{U}_1$ and $\widetilde{S}_1$ LQs, requiring the $\ell_{q} q \ell$ coupling to satisfy $\lesssim 0.3$ ensures that the contamination remains below $10$\% (provided $\ell_qq\tau_2$ couplings are order unity).

\section{Summary and Conclusions} 
\label{summary}
\noindent
In this work, we investigated novel pair production mechanisms of third-generation weak-singlet VLLs ($pp \to \tau_2\tau_2 + \text{jets}$) via LQs at the HL-LHC. Conventional VLL production at the LHC proceeds through QED interactions, resulting in relatively small cross sections. In contrast, if VLLs interact with LQs, the cross section for LQ-mediated VLL production can be significantly enhanced for larger couplings. Many UV-complete models predict the coexistence of LQs and VLLs, making our proposed search channels highly relevant from both LQ and VLL search perspectives. These channels not only open up new decay modes for LQs but also provide novel production mechanisms for VLLs.

We adopted a bottom-up approach by considering one LQ state and one VLL state, connected to a quark through a free coupling. Consequently, our framework involves three free parameters: the masses of the LQ and VLL, and the LQ–VLL–quark coupling. We have derived the discovery and exclusion sensitivities of this parameter space at the HL-LHC using both monolepton and dilepton final states. To achieve this, we performed a cut-based analysis with carefully optimised selection criteria to effectively separate the signal from SM backgrounds.

We found that the monolepton channel offers slightly better sensitivity than the dilepton channel. This enhancement arises because, in the monolepton channel, either the $\tau_2$ or one of the gauge bosons can decay hadronically, resulting in a larger overall cross section compared to the dilepton channel, which involves more leptonic branchings. In both cases, the impact of the optimised selection cuts is comparable. Overall, our analysis shows that the proposed correlated search for LQ and VLL signatures through these channels provides a promising and complementary search strategy for probing LQs and VLLs in future LHC runs.

\section*{acknowledgement}
\noindent S.D. and T.M. acknowledges partial support from the SERB/ANRF, Government of India, through the Core Research Grant No. CRG/2023/007031. N.K. would like to thank SERB/ANRF, Government of India for the support from the Startup Research Grant No. SRG/2022/000363 and Core Research Grant No. CRG/2022/004120. R.S. acknowledges the PMRF (ID: 0802000) from the Government of India.

\appendix
\section{Interaction Lagrangian} \label{app:tau2_gauge}

\subsection{Yukawa sector and mass matrix}
\noindent
The Yukawa Lagrangian before EWSB is 
\begin{align*}
    -\mathcal{L}_Y &= \lambda_{\tau}\,\overline{L}_{3L}H\tau_{R} 
    + \omega_{\tau}\,\overline{L}_{3L}H\tau'_{R} 
    + M_{\tau^{\prime}}\,\overline{\tau}'_{L}\tau'_{R} 
    + \text{h.c.}
    \label{eq:lagrangian_y}
\end{align*}
After EWSB, with $H \to (0,\, (v+h)/\sqrt{2})^T$, this becomes
\begin{align*}
    -\mathcal{L}_Y &= \frac{v\lambda_{\tau}}{\sqrt{2}}\overline{\tau}_{L}\tau_{R} 
    + \frac{v\omega_{\tau}}{\sqrt{2}}\overline{\tau}_{L}\tau'_{R} 
    + M_{\tau^{\prime}}\,\overline{\tau}'_{L}\tau'_{R} \notag \\
    &\quad + h\left(\frac{\lambda_{\tau}}{\sqrt{2}}\overline{\tau}_{L}\tau_{R} 
    + \frac{\omega_{\tau}}{\sqrt{2}}\overline{\tau}_{L}\tau'_{R}\right) 
    + \text{h.c.}
\end{align*}
This yields the following mass matrix with a non-diagonal mass term:
\begin{align*}
    \mathcal{L}_{\text{mass}} &= 
    -\begin{pmatrix}
      \overline{\tau}_{L} & \overline{\tau}'_{L}
    \end{pmatrix}
   \begin{pmatrix}
    \dfrac{\lambda_{\tau}v}{\sqrt{2}} & \dfrac{\omega_{\tau}v}{\sqrt{2}} \\
    0 & M_{\tau^{\prime}}
   \end{pmatrix}
   \begin{pmatrix}
      \tau_{R} \\ 
      \tau'_{R} 
   \end{pmatrix}
   + \text{h.c.}
\end{align*}
We define $M_{11} \equiv \lambda_{\tau}v/\sqrt{2}$, $M_{12} \equiv \omega_{\tau}v/\sqrt{2}$ and $M_{22} = M_{\tau^{\prime}}$.

\subsection{Mass diagonalisation and mixing angles}
\noindent
The physical states are obtained by diagonalising the mass matrix via a bi-unitary transformation. Denoting $s_{L,R} \equiv \sin\theta_{L,R}$ and $c_{L,R} \equiv \cos\theta_{L,R}$, the rotations are
\begin{align*}
\begin{pmatrix}
    \overline{\tau}_{X} & \overline{\tau}'_{X} 
\end{pmatrix}
&=
\begin{pmatrix}
    c_X & -s_X \\
    s_X & c_X 
\end{pmatrix}
\begin{pmatrix}
    \tau_{1X} \\ 
    \tau_{2X} 
\end{pmatrix}.
\end{align*}
Here $X = L/R$ and , $\{\tau_{1X}, \tau_{2X}\}$ are the mass eigenstates, while $\{\tau_{X}, \tau'_{X}\}$ denote flavour eigenstates.
The mixing angles are
\begin{align*}
\tan(2\theta_{L}) &= \frac{-2M_{12}M_{22}}{M_{22}^2 - M_{11}^2 - M_{12}^2}, \\
\tan(2\theta_{R}) &= \frac{-2M_{12}M_{11}}{M_{22}^2 - M_{11}^2 + M_{12}^2}.
\end{align*}
The corresponding squared mass eigenvalues are
\begin{align*}
    M_{\tau_{1,2}}^2 &= \frac{M_{22}^2}{2}\left[\left(1 + r_{11}^2 + r_{12}^2\right) 
    \mp \sqrt{\left(1 + r_{11}^2 + r_{12}^2\right)^2 - 4r_{11}^2}\right],
\end{align*}
where $r_{ij} \equiv M_{ij}/M_{22}$ (here $M_{21} = 0$).

\subsection{Interaction vertices}
\noindent
The interactions of VLL with $W$, $Z$ and $h$ bosons are given by
\paragraph{Higgs boson ($h$):}
\begin{align*}
    \mathcal{L}_H &= -\frac{1}{v}\Big[
    c_L(M_{11}c_R + M_{12}s_R)\overline{\tau}_{1L}\tau_{1R} \nonumber \\ 
    &\quad+ c_L(-M_{11}s_R + M_{12}c_R)\overline{\tau}_{1L}\tau_{2R} \nonumber \\
    &\quad - s_L(M_{11}c_R + M_{12}s_R)\overline{\tau}_{2L}\tau_{1R} \nonumber \\
    &\quad + s_L(M_{11}s_R - M_{12}c_R)\overline{\tau}_{2L}\tau_{2R}
    \Big]h + \text{h.c.}
\end{align*}
\paragraph{$Z$ boson:}
Mixing generates off-diagonal couplings, allowing decays such as $\tau_2 \to Z\tau_1$:
\begin{align*}
    \mathcal{L}_Z &= -\frac{g}{c_W}\Big[
    \left(s_W^2 - \frac{1}{2}c_L^2\right)\overline{\tau}_{1L}\gamma^\mu\tau_{1L} 
    + s_W^2\overline{\tau}_{1R}\gamma^\mu\tau_{1R} \\
    &\quad + \frac{1}{2}c_Ls_L(\overline{\tau}_{1L}\gamma^\mu\tau_{2L} + \overline{\tau}_{2L}\gamma^\mu\tau_{1L}) \\
    &\quad + \left(s_W^2 - \frac{1}{2}s_L^2\right)\overline{\tau}_{2L}\gamma^\mu\tau_{2L} 
    + s_W^2\overline{\tau}_{2R}\gamma^\mu\tau_{2R}\Big]Z_\mu.
\end{align*}

\paragraph{$W$ boson:}
The charged-current interaction arises solely from the $\tau_1$–$\tau_2$ mixing:
\begin{align*}
    \mathcal{L}_W &= -\frac{g}{\sqrt{2}}\left[c_L\,\overline{\nu}_L\gamma^\mu\tau_{1L} - s_L\,\overline{\nu}_L\gamma^\mu\tau_{2L}\right]W_\mu + \text{h.c.}.
\end{align*}

\subsection{Decay widths}
\noindent
The partial widths of the heavy state $\tau_2$ are:
\begin{align*}
\Gamma(\tau_2 \to W\nu) &= \frac{M_{\tau_2}^3}{32\pi M_W^2} 
 \left(\frac{g\,s_L}{\sqrt{2}}\right)^2\left(1 - \frac{M_W^2}{M_{\tau_2}^2}\right)^2
 \left(1 + \frac{2M_W^2}{M_{\tau_2}^2}\right), \\
\Gamma(\tau_2 \to Z\tau) &= \frac{M_{\tau_2}^3}{32\pi M_Z^2}\left(\frac{g\,c_Ls_L}{2c_W}\right)^2 
 \left(1 - \frac{M_Z^2}{M_{\tau_2}^2}\right)^2
 \left(1 + \frac{2M_Z^2}{M_{\tau_2}^2}\right), \\
\Gamma(\tau_2 \to h\tau) &= \frac{M_{\tau_2}}{32\pi} 
 \left(1 - \frac{M_H^2}{M_{\tau_2}^2}\right)^2 \nonumber\\
 &\quad 
 \Bigg[
   \frac{-2M_{\tau_{1}} M_{12}}{v^2}c_Rs_R(c_L^2 - s_L^2)\nonumber\\
   &\quad + \frac{M_{\tau_{1}}^2}{v^2}(c_L^2s_R^2 + c_R^2s_L^2)
   + \frac{M_{12}^2}{v^2}(c_L^2c_R^2 + s_L^2s_R^2)
 \Bigg].
\end{align*}

\bibliography{References}

\providecommand{\href}[2]{#2}\begingroup\raggedright\begin{thebibliography}{10}

\bibitem{Choudhury:2020cpm}
D.~Choudhury, K.~Deka, T.~Mandal and S.~Sadhukhan, \emph{{Neutrino and $Z'$
  phenomenology in an anomaly-free $\mathbf{U}(1)$ extension: role of
  higher-dimensional operators}},
  \href{http://dx.doi.org/10.1007/JHEP06(2020)111}{\emph{JHEP} {\bf 06} (2020)
  111}, [\href{http://arxiv.org/abs/2002.02349}{{\tt 2002.02349}}].

\bibitem{Deka:2021koh}
K.~Deka, T.~Mandal, A.~Mukherjee and S.~Sadhukhan, \emph{{Leptogenesis in an
  anomaly-free U(1) extension with higher-dimensional operators}},
  \href{http://dx.doi.org/10.1016/j.nuclphysb.2023.116213}{\emph{Nucl. Phys. B}
  {\bf 991} (2023) 116213}, [\href{http://arxiv.org/abs/2105.15088}{{\tt
  2105.15088}}].

\bibitem{ThomasArun:2021rwf}
M.~Thomas~Arun, T.~Mandal, S.~Mitra, A.~Mukherjee, L.~Priya and A.~Sampath,
  \emph{{Testing left-right symmetry with an inverse seesaw mechanism at the
  LHC}}, \href{http://dx.doi.org/10.1103/PhysRevD.105.115007}{\emph{Phys. Rev.
  D} {\bf 105} (2022) 115007}, [\href{http://arxiv.org/abs/2109.09585}{{\tt
  2109.09585}}].

\bibitem{Arun:2022ecj}
M.~T. Arun, A.~Chatterjee, T.~Mandal, S.~Mitra, A.~Mukherjee and K.~Nivedita,
  \emph{{Search for the Z' boson decaying to a right-handed neutrino pair in
  leptophobic U(1) models}},
  \href{http://dx.doi.org/10.1103/PhysRevD.106.095035}{\emph{Phys. Rev. D} {\bf
  106} (2022) 095035}, [\href{http://arxiv.org/abs/2204.02949}{{\tt
  2204.02949}}].

\bibitem{Bhaskar:2023xkm}
A.~Bhaskar, Y.~Chaurasia, K.~Deka, T.~Mandal, S.~Mitra and A.~Mukherjee,
  \emph{{Right-handed neutrino pair production via second-generation
  leptoquarks}},
  \href{http://dx.doi.org/10.1016/j.physletb.2023.138039}{\emph{Phys. Lett. B}
  {\bf 843} (2023) 138039}, [\href{http://arxiv.org/abs/2301.11889}{{\tt
  2301.11889}}].

\bibitem{Mandal:2023mck}
T.~Mandal, A.~Masaye, S.~Mitra, C.~Neeraj, N.~Reule and K.~Shah, \emph{{Pinning
  down the leptophobic $Z'$ in leptonic final states with Deep Learning}},
  \href{http://dx.doi.org/10.1016/j.physletb.2023.138417}{\emph{Phys. Lett. B}
  {\bf 849} (2024) 138417}, [\href{http://arxiv.org/abs/2307.01118}{{\tt
  2307.01118}}].

\bibitem{Duraikandan:2024kcy}
G.~Duraikandan, R.~Khanna, T.~Mandal, S.~Mitra and R.~Sharma,
  \emph{{Right-handed neutrino production through first-generation
  leptoquarks}},
  \href{http://dx.doi.org/10.1103/PhysRevD.111.075032}{\emph{Phys. Rev. D} {\bf
  111} (2025) 075032}, [\href{http://arxiv.org/abs/2412.19751}{{\tt
  2412.19751}}].

\bibitem{Poh:2017tfo}
Z.~Poh and S.~Raby, \emph{{Vectorlike leptons: Muon g-2 anomaly, lepton flavor
  violation, Higgs boson decays, and lepton nonuniversality}},
  \href{http://dx.doi.org/10.1103/PhysRevD.96.015032}{\emph{Phys. Rev. D} {\bf
  96} (2017) 015032}, [\href{http://arxiv.org/abs/1705.07007}{{\tt
  1705.07007}}].

\bibitem{Crivellin:2020ebi}
A.~Crivellin, F.~Kirk, C.~A. Manzari and M.~Montull, \emph{{Global Electroweak
  Fit and Vector-Like Leptons in Light of the Cabibbo Angle Anomaly}},
  \href{http://dx.doi.org/10.1007/JHEP12(2020)166}{\emph{JHEP} {\bf 12} (2020)
  166}, [\href{http://arxiv.org/abs/2008.01113}{{\tt 2008.01113}}].

\bibitem{Hamaguchi:2022byw}
K.~Hamaguchi, N.~Nagata, G.~Osaki and S.-Y. Tseng, \emph{{Probing new physics
  in the vector-like lepton model by lepton electric dipole moments}},
  \href{http://dx.doi.org/10.1007/JHEP01(2023)100}{\emph{JHEP} {\bf 01} (2023)
  100}, [\href{http://arxiv.org/abs/2211.16800}{{\tt 2211.16800}}].

\bibitem{delAguila:2008pw}
F.~del Aguila, J.~de~Blas and M.~Perez-Victoria, \emph{{Effects of new leptons
  in Electroweak Precision Data}},
  \href{http://dx.doi.org/10.1103/PhysRevD.78.013010}{\emph{Phys. Rev. D} {\bf
  78} (2008) 013010}, [\href{http://arxiv.org/abs/0803.4008}{{\tt 0803.4008}}].

\bibitem{Martin:2009bg}
S.~P. Martin, \emph{{Extra vector-like matter and the lightest Higgs scalar
  boson mass in low-energy supersymmetry}},
  \href{http://dx.doi.org/10.1103/PhysRevD.81.035004}{\emph{Phys. Rev. D} {\bf
  81} (2010) 035004}, [\href{http://arxiv.org/abs/0910.2732}{{\tt 0910.2732}}].

\bibitem{FileviezPerez:2011pt}
P.~Fileviez~Perez and M.~B. Wise, \emph{{Breaking Local Baryon and Lepton
  Number at the TeV Scale}},
  \href{http://dx.doi.org/10.1007/JHEP08(2011)068}{\emph{JHEP} {\bf 08} (2011)
  068}, [\href{http://arxiv.org/abs/1106.0343}{{\tt 1106.0343}}].

\bibitem{Joglekar:2012vc}
A.~Joglekar, P.~Schwaller and C.~E.~M. Wagner, \emph{{Dark Matter and Enhanced
  Higgs to Di-photon Rate from Vector-like Leptons}},
  \href{http://dx.doi.org/10.1007/JHEP12(2012)064}{\emph{JHEP} {\bf 12} (2012)
  064}, [\href{http://arxiv.org/abs/1207.4235}{{\tt 1207.4235}}].

\bibitem{Kearney:2012zi}
J.~Kearney, A.~Pierce and N.~Weiner, \emph{{Vectorlike Fermions and Higgs
  Couplings}}, \href{http://dx.doi.org/10.1103/PhysRevD.86.113005}{\emph{Phys.
  Rev. D} {\bf 86} (2012) 113005}, [\href{http://arxiv.org/abs/1207.7062}{{\tt
  1207.7062}}].

\bibitem{Kumar:2015tna}
N.~Kumar and S.~P. Martin, \emph{{Vectorlike Leptons at the Large Hadron
  Collider}}, \href{http://dx.doi.org/10.1103/PhysRevD.92.115018}{\emph{Phys.
  Rev. D} {\bf 92} (2015) 115018}, [\href{http://arxiv.org/abs/1510.03456}{{\tt
  1510.03456}}].

\bibitem{Bhattacharya:2018fus}
S.~Bhattacharya, P.~Ghosh, N.~Sahoo and N.~Sahu, \emph{{A Mini‑review on
  Vector‑like Leptonic Dark Matter, Neutrino Mass and Collider Signatures}},
  {\emph{Front. Phys.} {\bf 7} (2019) 80},
  [\href{http://arxiv.org/abs/1812.06505}{{\tt 1812.06505}}].

\bibitem{Chakraborty:2021tdo}
I.~Chakraborty, D.~K. Ghosh, N.~Ghosh and S.~K. Rai, \emph{{Dark Matter and
  Collider Searches in $S_3$-Symmetric 2HDM with Vector Like Lepton}},
  \href{http://dx.doi.org/10.1140/epjc/s10052-021-09446-5}{\emph{Eur. Phys. J.
  C} {\bf 81} (2021) 679}, [\href{http://arxiv.org/abs/2104.03351}{{\tt
  2104.03351}}].

\bibitem{Cherchiglia:2021syq}
A.~L. Cherchiglia, G.~De~Conto and C.~C. Nishi, \emph{{Leptonic CP violation
  from a vector-like lepton}},
  \href{http://dx.doi.org/10.1007/JHEP03(2022)010}{\emph{JHEP} {\bf 03} (2022)
  010}, [\href{http://arxiv.org/abs/2112.03943}{{\tt 2112.03943}}].

\bibitem{Bigaran:2023ris}
I.~Bigaran, B.~A. Dobrescu and A.~Russo, \emph{{Mutually elusive: Vectorlike
  antileptons and leptoquarks}},
  \href{http://dx.doi.org/10.1103/PhysRevD.109.055033}{\emph{Phys. Rev. D} {\bf
  109} (2024) 055033}, [\href{http://arxiv.org/abs/2312.09189}{{\tt
  2312.09189}}].

\bibitem{Cingiloglu:2024vdh}
K.~Y. Cingiloglu and M.~Frank, \emph{{Stability of the standard model vacuum
  with vectorlike leptons: A critical examination}},
  \href{http://dx.doi.org/10.1103/PhysRevD.111.016025}{\emph{Phys. Rev. D} {\bf
  111} (2025) 016025}, [\href{http://arxiv.org/abs/2408.10898}{{\tt
  2408.10898}}].

\bibitem{Kumar:2025aek}
R.~Kumar and R.~Srivastava, \emph{{Dark Matter Induced Proton Decays}},
  \href{http://arxiv.org/abs/2506.04370}{{\tt 2506.04370}}.

\bibitem{Bernreuther:2023uxh}
E.~Bernreuther and B.~A. Dobrescu, \emph{{Vectorlike leptons and long-lived
  bosons at the LHC}},
  \href{http://dx.doi.org/10.1007/JHEP07(2023)079}{\emph{JHEP} {\bf 07} (2023)
  079}, [\href{http://arxiv.org/abs/2304.08509}{{\tt 2304.08509}}].

\bibitem{Bandyopadhyay:2023joz}
P.~Bandyopadhyay, M.~Frank, S.~Parashar and C.~Sen, \emph{{Interplay of inert
  doublet and vector-like lepton triplet with displaced vertices at the LHC/FCC
  and MATHUSLA}}, \href{http://dx.doi.org/10.1007/JHEP03(2024)109}{\emph{JHEP}
  {\bf 03} (2024) 109}, [\href{http://arxiv.org/abs/2310.08883}{{\tt
  2310.08883}}].

\bibitem{Cao:2023smj}
Q.-H. Cao, J.~Guo, J.~Liu, Y.~Luo and X.-P. Wang, \emph{{Long-lived searches of
  vectorlike lepton and its accompanying scalar at colliders}},
  \href{http://dx.doi.org/10.1103/PhysRevD.110.015029}{\emph{Phys. Rev. D} {\bf
  110} (2024) 015029}, [\href{http://arxiv.org/abs/2311.12934}{{\tt
  2311.12934}}].

\bibitem{ATLAS:2024mrr}
{\bf ATLAS} collaboration, G.~Aad et~al., \emph{{Search for vector-like leptons
  coupling to first- and second-generation Standard Model leptons in pp
  collisions at $ \sqrt{s} $ = 13 TeV with the ATLAS detector}},
  \href{http://dx.doi.org/10.1007/JHEP05(2025)075}{\emph{JHEP} {\bf 05} (2025)
  075}, [\href{http://arxiv.org/abs/2411.07143}{{\tt 2411.07143}}].

\bibitem{CMS:2022nty}
{\bf CMS} collaboration, A.~Tumasyan et~al., \emph{{Inclusive nonresonant
  multilepton probes of new phenomena at $\sqrt s$=13\,\,TeV}},
  \href{http://dx.doi.org/10.1103/PhysRevD.105.112007}{\emph{Phys. Rev. D} {\bf
  105} (2022) 112007}, [\href{http://arxiv.org/abs/2202.08676}{{\tt
  2202.08676}}].

\bibitem{ATLAS:2023sbu}
{\bf ATLAS} collaboration, G.~Aad et~al., \emph{{Search for third-generation
  vector-like leptons in $pp$ collisions at $\sqrt{s} = 13\,\text{TeV}$ with
  the ATLAS detector}},
  \href{http://dx.doi.org/10.1007/JHEP07(2023)118}{\emph{JHEP} {\bf 07} (2023)
  118}, [\href{http://arxiv.org/abs/2303.05441}{{\tt 2303.05441}}].

\bibitem{Pati:1973uk}
J.~C. Pati and A.~Salam, \emph{{Unified Lepton-Hadron Symmetry and a Gauge
  Theory of the Basic Interactions}},
  \href{http://dx.doi.org/10.1103/PhysRevD.8.1240}{\emph{Phys. Rev. D} {\bf 8}
  (1973) 1240--1251}.

\bibitem{Pati:1974yy}
J.~C. Pati and A.~Salam, \emph{{Lepton Number as the Fourth Color}},
  \href{http://dx.doi.org/10.1103/PhysRevD.10.275}{\emph{Phys. Rev. D} {\bf 10}
  (1974) 275--289}. [Erratum: Phys.Rev.D 11, 703--703 (1975)].

\bibitem{Georgi:1974sy}
H.~Georgi and S.~L. Glashow, \emph{{Unity of All Elementary Particle Forces}},
  \href{http://dx.doi.org/10.1103/PhysRevLett.32.438}{\emph{Phys. Rev. Lett.}
  {\bf 32} (1974) 438--441}.

\bibitem{Fritzsch:1974nn}
H.~Fritzsch and P.~Minkowski, \emph{{Unified Interactions of Leptons and
  Hadrons}}, \href{http://dx.doi.org/10.1016/0003-4916(75)90211-0}{\emph{Annals
  Phys.} {\bf 93} (1975) 193--266}.

\bibitem{Schrempp:1984nj}
B.~Schrempp and F.~Schrempp, \emph{{LIGHT LEPTOQUARKS}},
  \href{http://dx.doi.org/10.1016/0370-2693(85)91450-9}{\emph{Phys. Lett. B}
  {\bf 153} (1985) 101--107}.

\bibitem{Kohda:2012sr}
M.~Kohda, H.~Sugiyama and K.~Tsumura, \emph{{Lepton number violation at the LHC
  with leptoquark and diquark}},
  \href{http://dx.doi.org/10.1016/j.physletb.2012.12.048}{\emph{Phys. Lett. B}
  {\bf 718} (2013) 1436--1440}, [\href{http://arxiv.org/abs/1210.5622}{{\tt
  1210.5622}}].

\bibitem{Dimopoulos:1979es}
S.~Dimopoulos and L.~Susskind, \emph{{Mass Without Scalars}},
  \href{http://dx.doi.org/10.1016/0550-3213(79)90364-X}{\emph{Nucl. Phys. B}
  {\bf 155} (1979) 237--252}.

\bibitem{Farhi:1980xs}
E.~Farhi and L.~Susskind, \emph{{Technicolor}},
  \href{http://dx.doi.org/10.1016/0370-1573(81)90173-3}{\emph{Phys. Rept.} {\bf
  74} (1981) 277}.

\bibitem{Barbier:2004ez}
R.~Barbier et~al., \emph{{R-parity violating supersymmetry}},
  \href{http://dx.doi.org/10.1016/j.physrep.2005.08.006}{\emph{Phys. Rept.}
  {\bf 420} (2005) 1--202}, [\href{http://arxiv.org/abs/hep-ph/0406039}{{\tt
  hep-ph/0406039}}].

\bibitem{Mandal:2015vfa}
T.~Mandal, S.~Mitra and S.~Seth, \emph{{Single Productions of Colored Particles
  at the LHC: An Example with Scalar Leptoquarks}},
  \href{http://dx.doi.org/10.1007/JHEP07(2015)028}{\emph{JHEP} {\bf 07} (2015)
  028}, [\href{http://arxiv.org/abs/1503.04689}{{\tt 1503.04689}}].

\bibitem{Mandal:2018kau}
T.~Mandal, S.~Mitra and S.~Raz, \emph{{$R_{D^{(*)}}$ motivated $S_1$ leptoquark
  scenarios: Impact of interference on the exclusion limits from LHC data}},
  \href{http://dx.doi.org/10.1103/PhysRevD.99.055028}{\emph{Phys. Rev. D} {\bf
  99} (2019) 055028}, [\href{http://arxiv.org/abs/1811.03561}{{\tt
  1811.03561}}].

\bibitem{Aydemir:2019ynb}
U.~Aydemir, T.~Mandal and S.~Mitra, \emph{{Addressing the $R_{D^{(*)}}$
  anomalies with an $S_1$ leptoquark from $SO(10)$ grand unification}},
  \href{http://dx.doi.org/10.1103/PhysRevD.101.015011}{\emph{Phys. Rev. D} {\bf
  101} (2020) 015011}, [\href{http://arxiv.org/abs/1902.08108}{{\tt
  1902.08108}}].

\bibitem{Chandak:2019iwj}
K.~Chandak, T.~Mandal and S.~Mitra, \emph{{Hunting for scalar leptoquarks with
  boosted tops and light leptons}},
  \href{http://dx.doi.org/10.1103/PhysRevD.100.075019}{\emph{Phys. Rev. D} {\bf
  100} (2019) 075019}, [\href{http://arxiv.org/abs/1907.11194}{{\tt
  1907.11194}}].

\bibitem{Bhaskar:2020kdr}
A.~Bhaskar, D.~Das, B.~De and S.~Mitra, \emph{{Enhancing scalar productions
  with leptoquarks at the LHC}},
  \href{http://dx.doi.org/10.1103/PhysRevD.102.035002}{\emph{Phys. Rev. D} {\bf
  102} (2020) 035002}, [\href{http://arxiv.org/abs/2002.12571}{{\tt
  2002.12571}}].

\bibitem{Bhaskar:2020gkk}
A.~Bhaskar, T.~Mandal and S.~Mitra, \emph{{Boosting vector leptoquark searches
  with boosted tops}},
  \href{http://dx.doi.org/10.1103/PhysRevD.101.115015}{\emph{Phys. Rev. D} {\bf
  101} (2020) 115015}, [\href{http://arxiv.org/abs/2004.01096}{{\tt
  2004.01096}}].

\bibitem{Bhaskar:2021pml}
A.~Bhaskar, D.~Das, T.~Mandal, S.~Mitra and C.~Neeraj, \emph{{Precise limits on
  the charge-$2/3$ $U_1$ vector leptoquark}},
  \href{http://dx.doi.org/10.1103/PhysRevD.104.035016}{\emph{Phys. Rev. D} {\bf
  104} (2021) 035016}, [\href{http://arxiv.org/abs/2101.12069}{{\tt
  2101.12069}}].

\bibitem{Bhaskar:2021gsy}
A.~Bhaskar, T.~Mandal, S.~Mitra and M.~Sharma, \emph{{Improving
  third-generation leptoquark searches with combined signals and boosted top
  quarks}}, \href{http://dx.doi.org/10.1103/PhysRevD.104.075037}{\emph{Phys.
  Rev. D} {\bf 104} (2021) 075037},
  [\href{http://arxiv.org/abs/2106.07605}{{\tt 2106.07605}}].

\bibitem{Bandyopadhyay:2021pld}
P.~Bandyopadhyay, A.~Karan, R.~Mandal and S.~Parashar, \emph{{Distinguishing
  signatures of scalar leptoquarks at hadron and muon colliders}},
  \href{http://dx.doi.org/10.1140/epjc/s10052-022-10809-9}{\emph{Eur. Phys. J.
  C} {\bf 82} (2022) 916}, [\href{http://arxiv.org/abs/2108.06506}{{\tt
  2108.06506}}].

\bibitem{Bhaskar:2022vgk}
A.~Bhaskar, A.~A. Madathil, T.~Mandal and S.~Mitra, \emph{{Combined explanation
  of W-mass, muon g-2, RK(*) and RD(*) anomalies in a singlet-triplet scalar
  leptoquark model}},
  \href{http://dx.doi.org/10.1103/PhysRevD.106.115009}{\emph{Phys. Rev. D} {\bf
  106} (2022) 115009}, [\href{http://arxiv.org/abs/2204.09031}{{\tt
  2204.09031}}].

\bibitem{Aydemir:2022lrq}
U.~Aydemir, T.~Mandal, S.~Mitra and S.~Munir, \emph{{An economical model for
  $B$-flavour and $a_\mu$ anomalies from SO(10) grand unification}},
  \href{http://arxiv.org/abs/2209.04705}{{\tt 2209.04705}}.

\bibitem{Bhaskar:2023ftn}
A.~Bhaskar, A.~Das, T.~Mandal, S.~Mitra and R.~Sharma, \emph{{Fresh look at the
  LHC limits on scalar leptoquarks}},
  \href{http://dx.doi.org/10.1103/PhysRevD.109.055018}{\emph{Phys. Rev. D} {\bf
  109} (2024) 055018}, [\href{http://arxiv.org/abs/2312.09855}{{\tt
  2312.09855}}].

\bibitem{Cheung:2023gwm}
K.~Cheung, T.~T.~Q. Nguyen and C.~J. Ouseph, \emph{{Leptoquark search at the
  Forward Physics Facility}},
  \href{http://dx.doi.org/10.1103/PhysRevD.108.036014}{\emph{Phys. Rev. D} {\bf
  108} (2023) 036014}, [\href{http://arxiv.org/abs/2302.05461}{{\tt
  2302.05461}}].

\bibitem{Bhaskar:2024swq}
A.~Bhaskar, D.~Das, S.~Kundu, A.~A.~Madathil, T.~Mandal and S.~Mitra,
  \emph{{Vector leptoquark contributions to lepton dipole moments}},
  \href{http://dx.doi.org/10.1103/PhysRevD.111.015045}{\emph{Phys. Rev. D} {\bf
  111} (2025) 015045}, [\href{http://arxiv.org/abs/2408.11798}{{\tt
  2408.11798}}].

\bibitem{Bhaskar:2024snl}
A.~Bhaskar and M.~Mitra, \emph{{Boosted top quark inspired leptoquark searches
  at the muon collider}},  \href{http://arxiv.org/abs/2409.15992}{{\tt
  2409.15992}}.

\bibitem{Bhaskar:2024wic}
A.~Bhaskar, Y.~Chaurasia, A.~Das, A.~Kumar, T.~Mandal, S.~Mitra et~al.,
  \emph{{TooLQit: Leptoquark Models and Limits}},
  \href{http://arxiv.org/abs/2412.19729}{{\tt 2412.19729}}.

\bibitem{Das:2025osr}
A.~Das, T.~Mandal, S.~Mitra and R.~Sharma, \emph{{Fresh look at the LHC limits
  on vector leptoquarks}},  \href{http://arxiv.org/abs/2507.18295}{{\tt
  2507.18295}}.

\bibitem{ATL-PHYS-PUB-2024-012}
{\bf ATLAS} collaboration, \emph{{Leptoquark summary plot for scalar or vector
  models}},   , CERN, Geneva, 2024.

\bibitem{CMSPlot}
``{CMS Exotica Summary plots for 13 TeV data: Leptoquark summary plot}.''
  \url{https://twiki.cern.ch/twiki/pub/CMSPublic/SummaryPlotsEXO13TeV/barplot_QE_QMU_QTAU_QNU_2025March.pdf},
  2025.

\bibitem{DiLuzio:2017vat}
L.~Di~Luzio, A.~Greljo and M.~Nardecchia, \emph{{Gauge leptoquark as the origin
  of B-physics anomalies}},
  \href{http://dx.doi.org/10.1103/PhysRevD.96.115011}{\emph{Phys. Rev. D} {\bf
  96} (2017) 115011}, [\href{http://arxiv.org/abs/1708.08450}{{\tt
  1708.08450}}].

\bibitem{Calibbi:2017qbu}
L.~Calibbi, A.~Crivellin and T.~Li, \emph{{Model of vector leptoquarks in view
  of the $B$-physics anomalies}},
  \href{http://dx.doi.org/10.1103/PhysRevD.98.115002}{\emph{Phys. Rev. D} {\bf
  98} (2018) 115002}, [\href{http://arxiv.org/abs/1709.00692}{{\tt
  1709.00692}}].

\bibitem{DiLuzio:2018zxy}
L.~Di~Luzio, J.~Fuentes-Martin, A.~Greljo, M.~Nardecchia and S.~Renner,
  \emph{{Maximal Flavour Violation: a Cabibbo mechanism for leptoquarks}},
  \href{http://dx.doi.org/10.1007/JHEP11(2018)081}{\emph{JHEP} {\bf 11} (2018)
  081}, [\href{http://arxiv.org/abs/1808.00942}{{\tt 1808.00942}}].

\bibitem{CMS:2022cpe}
{\bf CMS} collaboration, A.~Tumasyan et~al., \emph{{Search for pair-produced
  vector-like leptons in final states with third-generation leptons and at
  least three b quark jets in proton-proton collisions at $\sqrt{s} =13$ TeV}},
  \href{http://dx.doi.org/10.1016/j.physletb.2023.137713}{\emph{Phys. Lett. B}
  {\bf 846} (2023) 137713}, [\href{http://arxiv.org/abs/2208.09700}{{\tt
  2208.09700}}].

\bibitem{Bhardwaj:2022nko}
A.~Bhardwaj, T.~Mandal, S.~Mitra and C.~Neeraj, \emph{{Roadmap to explore
  vectorlike quarks decaying to a new scalar or pseudoscalar}},
  \href{http://dx.doi.org/10.1103/PhysRevD.106.095014}{\emph{Phys. Rev. D} {\bf
  106} (2022) 095014}, [\href{http://arxiv.org/abs/2203.13753}{{\tt
  2203.13753}}].

\bibitem{Blumlein:1994tu}
J.~Bl\"umlein and E.~Boos, \emph{{Leptoquark production at high energy $e^{+}
  e^{-}$ colliders}},
  \href{http://dx.doi.org/10.1016/0920-5632(94)90675-0}{\emph{Nucl. Phys. B
  Proc. Suppl.} {\bf 37} (1994) 181--192}.

\bibitem{Blumlein:1996qp}
J.~Blumlein, E.~Boos and A.~Kryukov, \emph{{Leptoquark pair production in
  hadronic interactions}},
  \href{http://dx.doi.org/10.1007/s002880050538}{\emph{Z. Phys. C} {\bf 76}
  (1997) 137--153}, [\href{http://arxiv.org/abs/hep-ph/9610408}{{\tt
  hep-ph/9610408}}].

\bibitem{Alloul:2013bka}
A.~Alloul, N.~D. Christensen, C.~Degrande, C.~Duhr and B.~Fuks,
  \emph{{FeynRules 2.0 - A complete toolbox for tree-level phenomenology}},
  \href{http://dx.doi.org/10.1016/j.cpc.2014.04.012}{\emph{Comput. Phys.
  Commun.} {\bf 185} (2014) 2250--2300},
  [\href{http://arxiv.org/abs/1310.1921}{{\tt 1310.1921}}].

\bibitem{Alwall:2014hca}
J.~Alwall, R.~Frederix, S.~Frixione, V.~Hirschi, F.~Maltoni, O.~Mattelaer
  et~al., \emph{{The automated computation of tree-level and next-to-leading
  order differential cross sections, and their matching to parton shower
  simulations}}, \href{http://dx.doi.org/10.1007/JHEP07(2014)079}{\emph{JHEP}
  {\bf 07} (2014) 079}, [\href{http://arxiv.org/abs/1405.0301}{{\tt
  1405.0301}}].

\bibitem{NNPDF:2021uiq}
{\bf NNPDF} collaboration, R.~D. Ball et~al., \emph{{An open-source machine
  learning framework for global analyses of parton distributions}},
  \href{http://dx.doi.org/10.1140/epjc/s10052-021-09747-9}{\emph{Eur. Phys. J.
  C} {\bf 81} (2021) 958}, [\href{http://arxiv.org/abs/2109.02671}{{\tt
  2109.02671}}].

\bibitem{Kramer:2004df}
M.~Kramer, T.~Plehn, M.~Spira and P.~M. Zerwas, \emph{{Pair production of
  scalar leptoquarks at the CERN LHC}},
  \href{http://dx.doi.org/10.1103/PhysRevD.71.057503}{\emph{Phys. Rev. D} {\bf
  71} (2005) 057503}, [\href{http://arxiv.org/abs/hep-ph/0411038}{{\tt
  hep-ph/0411038}}].

\bibitem{Mandal:2015lca}
T.~Mandal, S.~Mitra and S.~Seth, \emph{{Pair Production of Scalar Leptoquarks
  at the LHC to NLO Parton Shower Accuracy}},
  \href{http://dx.doi.org/10.1103/PhysRevD.93.035018}{\emph{Phys. Rev. D} {\bf
  93} (2016) 035018}, [\href{http://arxiv.org/abs/1506.07369}{{\tt
  1506.07369}}].

\bibitem{Borschensky:2020hot}
C.~Borschensky, B.~Fuks, A.~Kulesza and D.~Schwartl\"ander, \emph{{Scalar
  leptoquark pair production at hadron colliders}},
  \href{http://dx.doi.org/10.1103/PhysRevD.101.115017}{\emph{Phys. Rev. D} {\bf
  101} (2020) 115017}, [\href{http://arxiv.org/abs/2002.08971}{{\tt
  2002.08971}}].

\bibitem{Borschensky:2021hbo}
C.~Borschensky, B.~Fuks, A.~Kulesza and D.~Schwartl\"ander, \emph{{Scalar
  leptoquark pair production at the LHC: precision predictions in the era of
  flavour anomalies}},
  \href{http://dx.doi.org/10.1007/JHEP02(2022)157}{\emph{JHEP} {\bf 02} (2022)
  157}, [\href{http://arxiv.org/abs/2108.11404}{{\tt 2108.11404}}].

\bibitem{Borschensky:2022xsa}
C.~Borschensky, B.~Fuks, A.~Jueid and A.~Kulesza, \emph{{Scalar leptoquarks at
  the LHC and flavour anomalies: a comparison of pair-production modes at
  NLO-QCD}}, \href{http://dx.doi.org/10.1007/JHEP11(2022)006}{\emph{JHEP} {\bf
  11} (2022) 006}, [\href{http://arxiv.org/abs/2207.02879}{{\tt 2207.02879}}].

\bibitem{Bierlich:2022pfr}
C.~Bierlich et~al., \emph{{A comprehensive guide to the physics and usage of
  PYTHIA 8.3}},
  \href{http://dx.doi.org/10.21468/SciPostPhysCodeb.8}{\emph{SciPost Phys.
  Codeb.} {\bf 2022} (2022) 8}, [\href{http://arxiv.org/abs/2203.11601}{{\tt
  2203.11601}}].

\bibitem{deFavereau:2013fsa}
{\bf DELPHES 3} collaboration, J.~de~Favereau, C.~Delaere, P.~Demin,
  A.~Giammanco, V.~Lema\^\i{}tre, A.~Mertens et~al., \emph{{DELPHES 3, A
  modular framework for fast simulation of a generic collider experiment}},
  \href{http://dx.doi.org/10.1007/JHEP02(2014)057}{\emph{JHEP} {\bf 02} (2014)
  057}, [\href{http://arxiv.org/abs/1307.6346}{{\tt 1307.6346}}].

\bibitem{Cacciari:2011ma}
M.~Cacciari, G.~P. Salam and G.~Soyez, \emph{{FastJet User Manual}},
  \href{http://dx.doi.org/10.1140/epjc/s10052-012-1896-2}{\emph{Eur. Phys. J.
  C} {\bf 72} (2012) 1896}, [\href{http://arxiv.org/abs/1111.6097}{{\tt
  1111.6097}}].

\bibitem{Cacciari:2008gp}
M.~Cacciari, G.~P. Salam and G.~Soyez, \emph{{The anti-$k_t$ jet clustering
  algorithm}},
  \href{http://dx.doi.org/10.1088/1126-6708/2008/04/063}{\emph{JHEP} {\bf 04}
  (2008) 063}, [\href{http://arxiv.org/abs/0802.1189}{{\tt 0802.1189}}].

\bibitem{Catani:2009sm}
S.~Catani, L.~Cieri, G.~Ferrera, D.~de~Florian and M.~Grazzini, \emph{Vector
  boson production at hadron colliders: A fully exclusive qcd calculation at
  next-to-next-to-leading order},
  \href{http://dx.doi.org/10.1103/PhysRevLett.103.082001}{\emph{Phys. Rev.
  Lett.} {\bf 103} (Aug, 2009) 082001}.

\bibitem{Balossini:2009sa}
G.~Balossini, G.~Montagna, C.~M. Carloni~Calame, M.~Moretti, O.~Nicrosini,
  F.~Piccinini et~al., \emph{Combination of electroweak and qcd corrections to
  single w production at the fermilab tevatron and the cern lhc},
  \href{http://dx.doi.org/10.1007/jhep01(2010)013}{\emph{Journal of High Energy
  Physics} {\bf 2010} (Jan., 2010) }.

\bibitem{Campbell:2011bn}
J.~M. Campbell, R.~K. Ellis and C.~Williams, \emph{Vector boson pair production
  at the lhc}, \href{http://dx.doi.org/10.1007/jhep07(2011)018}{\emph{Journal
  of High Energy Physics} {\bf 2011} (July, 2011) }.

\bibitem{Kidonakis:2015nna}
N.~Kidonakis, \emph{{Theoretical results for electroweak-boson and single-top
  production}}, \href{http://dx.doi.org/10.22323/1.247.0170}{\emph{PoS} {\bf
  DIS2015} (2015) 170}, [\href{http://arxiv.org/abs/1506.04072}{{\tt
  1506.04072}}].

\bibitem{Muselli:2015kba}
C.~Muselli, M.~Bonvini, S.~Forte, S.~Marzani and G.~Ridolfi, \emph{{Top Quark
  Pair Production beyond NNLO}},
  \href{http://dx.doi.org/10.1007/JHEP08(2015)076}{\emph{JHEP} {\bf 08} (2015)
  076}, [\href{http://arxiv.org/abs/1505.02006}{{\tt 1505.02006}}].

\bibitem{Kulesza:2018tqz}
A.~Kulesza, L.~Motyka, D.~Schwartl\"ander, T.~Stebel and V.~Theeuwes,
  \emph{{Associated production of a top quark pair with a heavy electroweak
  gauge boson at NLO$+$NNLL accuracy}},
  \href{http://dx.doi.org/10.1140/epjc/s10052-019-6746-z}{\emph{Eur. Phys. J.
  C} {\bf 79} (2019) 249}, [\href{http://arxiv.org/abs/1812.08622}{{\tt
  1812.08622}}].

\bibitem{Kang16}
Z.~Kang, P.~Ko and J.~Li, \emph{New avenues to heavy right-handed neutrinos
  with pair production at hadronic colliders},
  \href{http://dx.doi.org/10.1103/PhysRevD.93.075037}{\emph{Phys. Rev. D} {\bf
  93} (Apr, 2016) 075037}.

\bibitem{Accomando:2017qcs}
E.~Accomando, L.~Delle~Rose, S.~Moretti, E.~Olaiya and C.~H.
  Shepherd-Themistocleous, \emph{{Extra Higgs boson and $Z^{\prime}$ as portals
  to signatures of heavy neutrinos at the LHC}},
  \href{http://dx.doi.org/10.1007/JHEP02(2018)109}{\emph{JHEP} {\bf 02} (2018)
  109}, [\href{http://arxiv.org/abs/1708.03650}{{\tt 1708.03650}}].

\bibitem{Helo:2018rll}
J.~C. Helo, H.~Li, N.~A. Neill, M.~Ramsey-Musolf and J.~C. Vasquez,
  \emph{{Probing neutrino Dirac mass in left-right symmetric models at the LHC
  and next generation colliders}},
  \href{http://dx.doi.org/10.1103/PhysRevD.99.055042}{\emph{Phys. Rev. D} {\bf
  99} (2019) 055042}, [\href{http://arxiv.org/abs/1812.01630}{{\tt
  1812.01630}}].

\bibitem{Huitu08}
K.~Huitu, S.~Khalil, H.~Okada and S.~K. Rai, \emph{Signatures for right-handed
  neutrinos at the large hadron collider},
  \href{http://dx.doi.org/10.1103/PhysRevLett.101.181802}{\emph{Phys. Rev.
  Lett.} {\bf 101} (Oct, 2008) 181802}.

\bibitem{Bhardwaj:2019mts}
A.~Bhardwaj, P.~Konar, T.~Mandal and S.~Sadhukhan, \emph{{Probing the inert
  doublet model using jet substructure with a multivariate analysis}},
  \href{http://dx.doi.org/10.1103/PhysRevD.100.055040}{\emph{Phys. Rev. D} {\bf
  100} (2019) 055040}, [\href{http://arxiv.org/abs/1905.04195}{{\tt
  1905.04195}}].

\bibitem{Cowan:2010js}
G.~Cowan, K.~Cranmer, E.~Gross and O.~Vitells, \emph{{Asymptotic formulae for
  likelihood-based tests of new physics}},
  \href{http://dx.doi.org/10.1140/epjc/s10052-011-1554-0}{\emph{Eur. Phys. J.
  C} {\bf 71} (2011) 1554}, [\href{http://arxiv.org/abs/1007.1727}{{\tt
  1007.1727}}]. [Erratum: Eur.Phys.J.C 73, 2501 (2013)].

\end{thebibliography}\endgroup
\bibliographystyle{JHEPCust}

\end{document}